\documentclass[10pt]{article}
\pdfoutput=1
\usepackage{amstext,enumerate,amssymb,amsmath,bbm}
\usepackage{esint}
\usepackage{slashed}
\usepackage{framed}
\usepackage{setspace}
\usepackage{cite}
\usepackage{float}
\usepackage{multirow}
\usepackage{enumitem}
\usepackage{graphicx}
  
\newcommand{\n}{\nabla}

\newcommand{\z}{\zeta}

\newcommand{\cA}{\mathcal{A}}
\newcommand{\cH}{\mathcal{H}}

\newcommand{\cM}{\mathcal{M}}
\newcommand{\cU}{\mathcal{U}}

\makeatletter
\renewcommand\paragraph{\@startsection{paragraph}{4}{\z@}%
            {-2.5ex\@plus -1ex \@minus -.25ex}%
            {1.25ex \@plus .25ex}%
            {\normalfont\normalsize\bfseries}}
\makeatother
\numberwithin{equation}{section}

\begin{document}

 \rightline{LTH 1099}
\rightline{DMUS--MP--16/19}
\vskip 1.5 true cm  
\begin{center}  
{\large Five-dimensional Nernst branes from special geometry}\\[.5em]
\vskip 1.0 true cm   
{P.~Dempster,$^{1}$ D.~Errington,$^1$ J.~Gutowski$^2$ and T.~Mohaupt$^{1}$} \\[5pt] 
$^1${Department of Mathematical Sciences\\ 
University of Liverpool\\
Peach Street \\
Liverpool L69 7ZL, UK\\  
pdempster2@gmail.com,
D.Errington@liv.ac.uk, Thomas.Mohaupt@liv.ac.uk \\[1em]
}
$^2${Department of Mathematics\\
University of Surrey\\
Guildford, GU2 7XH, UK\\
j.gutowski@surrey.ac.uk
}\\[2ex]
September 16, 2016. Revised: October 11, 2016. Final version: 
November 15, 2016.
\end{center}  
\vskip 1.0 true cm  
\baselineskip=18pt  
\begin{abstract}  
\noindent  
We construct Nernst brane solutions, that is black branes with
zero entropy density in the extremal limit, of FI-gauged minimal
five-dimensional supergravity coupled to an arbitrary number
of vector multiplets. While the scalars take specific constant values and
dynamically determine the value of the cosmological constant
in terms of the FI-parameters, the metric takes the 
form of a boosted AdS Schwarzschild black brane. This metric can 
be brought to the Carter-Novotn\'y-Horsk\'y form that has previously
been observed to occur in certain limits of boosted D3-branes. By
dimensional reduction to four dimensions we recover the four-dimensional
Nernst branes of arXiv:1501.07863 and show how the five-dimensional
lift resolves all their UV singularities. The dynamics of the 
compactification circle, which expands both in the UV and in the 
IR, plays a crucial role. At asymptotic infinity, the curvature singularity 
of the four-dimensional metric and the run-away behaviour of the
four-dimensional scalar combine in such a way that the lifted solution
becomes asymptotic to AdS$_5$. Moreover, the existence of a finite
chemical potential in four dimensions is related to fact that the
compactification circle has a finite minimal value. While it is not
clear immediately how to embed our solutions into string theory, we
argue that the same type of dictionary as proposed for boosted D3-branes
should apply, although with a lower amount of supersymmetry.

\end{abstract}


\newpage
 
\tableofcontents

\section{Introduction}

The Nernst law or third law of thermodynamics comes in two versions.
The weak version, which states that zero temperature can only be reached
asymptotically, is uncontroversial. In constrast, there is an ongoing
discussion about the status of the strong version, originally formulated
by Planck, which states that entropy goes to zero at zero temperature.
See for example 
\cite{Wald:1997qp,Hartnoll:2009sz,DHoker:2009bc,Hartnoll:2011fn} for 
contrasting views on this. With regard to 
gauge/gravity duality, there is a natural tension between condensed 
matter systems, where the strong version is believed to apply generally
or at least generically, and BPS and other extremal black hole solutions,
where a regular, and hence normally finite, horizon is associated with a
finite, and typically large entropy. This raises the question whether and
how systems obyeing the strong version of the Nernst law can be modelled
by gravitational counterparts.

Extremal black brane solutions obeying the strong version of the 
Nernst law have been found in a variety of theories \cite{DHoker:2009bc,Ammon:2016szz,Goldstein:2009cv,Goldstein:2010aw,Gauntlett:2009dn,Horowitz:2009ij}, including
four-dimensional FI-gauged supergravity 
\cite{Barisch:2011ui}, where they were
dubbed Nernst branes. More recently, 
a two-parameter family of Nernst branes parametrized by temperature $T$
and a chemical potential $\mu$ was found in \cite{Dempster:2015xqa}. 
Asymptotically, these solutions approach hyperscaling violating 
Lifshitz (`hvLif') geometries both at infinity and at the horizon, and therefore
are interesting in the context of gauge/gravity duality with 
hyperscaling violation \cite{Huijse:2011ef,Dong:2012se}, see
also \cite{Perlmutter:2012he} and references therein. Four-dimensional
Nernst branes share the typical problems of hvLif geometries,
in that they exhibit curvature singularities 
\cite{Horowitz:2011gh,Copsey:2012gw}. Moreover the scaling properties
of the geometry at infinity suggested an entropy--temperature relation
of the form $S\sim T^3$, while the high-temperature asymptotics 
extracted from the equation of state was found to be $S\sim T$.\footnote{
Since the brane world volume is infinite, extensive quantities such 
as entropy are supposed to be taken per unit volume.} 
Since in addition the scalars showed runaway behaviour at infinity,
and the relation $S\sim T^3$ is valid for AdS$_5$, 
it was conjectured in \cite{Dempster:2015xqa} that the inconsistent
UV behaviour of four-dimensional Nernst branes signals a dynamical
decompactification, and that the above problem would be cured by
lifting the solutions to five dimensions. 
This follows the general idea that scale covariant vacua can be
obtained by dimensional reduction of scale invariant vacua 
\cite{Perlmutter:2012he}.

In this paper we will verify this proposal and study the relation 
between five-dimensional and four-dimensional Nernst branes in 
detail. The five-dimensional Nernst branes will be constructed
within FI-gauged minimal five-dimensional supergravity with an 
arbitrary number of vector multiplets, by dimensional
reduction to an effective three-dimensional Euclidean theory 
and using the special geometry of the scalar sector. We will show that 
the singular asymptotic behaviour of 
four-dimensional Nernst branes is a compactification artefact, and that
the ground state geometry is AdS$_5$. A crucial role in understanding
the relation between five- and four-dimensional Nernst branes is played
by the compactification circle, whose size changes dynamically along
the direction transverse to the brane. The behaviour of this circle
also solves another puzzle, namely the origin of the four-dimensional
chemical potential. Five-dimensional Nernst branes turn out to 
be boosted AdS Schwarzschild branes, depending on two continuous
parameters, the temperature $T$, and the linear momentum $P_z$. Since momentum
turns into electric charge upon dimensional reduction, one might naively
expect that four-dimensional Nernst branes depend on one continuous parameter,
temperature $T$ and on one discrete parameter, electric charge $Q_0$. 
However, the solutions of \cite{Dempster:2015xqa} depend on an additional
continuous parameter, the chemical potential $\mu$. As it turns out,
its origin can be traced to the fact that the compactification 
circle grows towards infinity and towards the horizon, and has a minimum
in between. This minimum introduces a new scale, and since the minimal
value of the radius can be varied continuously, this provides an additional
continuous parameter.

The boosted AdS Schwarzschild metric we obtain by solving the
five-dimensional equations of motion is an Einstein metric and
can be brought to Carter-Novotn\'y-Horsk\'y form. Such metrics
describe the
near horizon regions of dimensionally reduced D-branes and M-branes
with superimposed pp-waves \cite{Cvetic:1998jf}. This does, however,
not immediately provide us with a string theory embedding of
our solutions, unless we switch off the vector multiplets. The solution
we find is valid for an arbitrary number of vector multiplets, and
depends on the choice of the prepotential and on the choice of an
FI-gauging through parameters $c_{ijk}$ and $g_i$. While the scalars
are constant, they still have to extremize the scalar potential, and
therefore these parameters determine the effective cosmological 
constant, and enter into the various integration constants of our
solution. We will give explicit expressions in the paper. 
FI-gauged five-dimensional $N=2$ supergravity\footnote{We 
count supersymmetry in four-dimensional units.} has so far only been
obtained as a consistent truncation of a higher-dimensional
supergravity in a very limited number of cases. 
The case without vector multiplets, that is
pure gauged five-dimensional $N=2$ supergravity,
can be obtained by reduction of IIB supergravity on
Sasaki-Einstein manifolds $Y^{p,q}$ \cite{Buchel:2006gb}. 
The STU-model and consistent truncations thereof
can be obtained as consistent reductions of eleven-dimensional
supergravity \cite{Cvetic:2000tb,Azizi:2016noi}.
Other consistent
truncations involve hypermultiplets or massive vector multiplets
and consequently have different types of gauging \cite{Cassani:2010uw,Gauntlett:2010vu}. 
The dimensionally
reduced boosted D3-branes of \cite{Cvetic:1998jf}
which lead to the same five-dimensional
metric should be considered as solutions of five-dimensional 
gauged $N=8$ supergravity,
which can be obtained by reduction of IIB supergravity on $S^5$. 
In this case the five-dimensional cosmological constant is simply 
determined by the D3-charge, and we cannot account for the
parameters $c_{ijk}, g_i$ of an FI-gauged supergravity theory with
vector multiplets. But while there is no obvious string theory 
embedding of our solutions, the five-dimensional metric is still the
same as for boosted D3-branes. Therefore it is 
reasonable to assume that at least the same type of dictionary
between geometry and field theory will apply. We will come back 
to this in the conclusions. Throughout the paper we keep a strictly
five-dimensional perspective and work with relations between 
geometric and thermodynamical quantities without using any higher-dimensional
or stringy input.

Our work includes a detailed study of the thermodynamical properties
of five-dimensional Nernst branes. Using the quasilocal energy momentum tensor
we construct expressions for the mass $M$ and momentum $P_z$. From the
near horizon behaviour of the solution we obtain the entropy $S$ and
through the surface gravity of the Killing horizon, the temperature $T$. 
We verify the validity of the first law, as well as the strong
version of the third law. The solution is shown to be thermodynamically 
stable. We obtain an equation of state and show 
that the relation between entropy and temperature interpolates
between $S\sim T^3$ at high temperature and $S\sim T^{1/3}$ at low
temparture. This asymptotic behaviour agrees with the literature on 
boosted D3-branes\cite{Cvetic:1998jf,Singh:2010zs} 
and verifies
the prediction of \cite{Dempster:2015xqa}. One subtlety is that
the metric admits a reparametrisation, which naively removes the 
integration constant corresponding to the temperature (as long as temperature
is non-zero) from the solution.\footnote{A related observation was made in 
\cite{Cvetic:1998jf}, where it was pointed out that one can locally 
remove the pp wave from the non-extremal 
solution by a coordinate transformation.} 
However, as the detailed analysis shows, 
when properly setting up thermodynamics using the quasi-local stress 
energy tensor, temperature is defined by a reparametrisation invariant
expression. The additional input that thermodynamics requires is the 
choice of the norm of the static Killing vector field, which should be
considered as part of the choice of the AdS$_5$ groundstate.

Besides the trivial
extremal limit, which is global AdS$_5$, the solution admits a non-trivial
double scaling limit, where temperature goes to zero, and the boost 
parameter goes to infinity while the momentum (density) is kept 
fixed. This limit
was studied (in different coordinates) in \cite{Cvetic:1998jf}, where it 
was shown to result in a homogenoeus Einstein space of Kaigorodov type, which 
is 1/4 BPS. This analysis applies to our solution and implies that 
it supports 2 out of a maximum
of 8 Killing spinors. In the extremal limit we also recover the
five-dimensional extremal Nernst branes of \cite{BarischDick:2012gj}.
There are interesting parallels as well as
differences between boosted AdS Schwarzschild black branes and rotating
black holes. Like for a Kerr black hole, boosted branes have an
`ergoregion' that is a region before the event horizon where observers
cannot stay static any more, but have to co-translate with the brane. 
Also, the Euclidean continuation of such a brane looks very similar
to that of a Kerr black hole, and allows to derive the temperature by imposing
the absence of a conical deficit. In other aspects the analogy breaks
down, however. While supersymmetric rotating black holes cannot
have an ergosphere \cite{Cvetic:2005zi}, the ergoregion of an 
infinitely boosted black brane remains. We show that this is consistent with 
supersymmetry, because the Killing vector obtained as a Killing spinor
bilinear is null rather than timelike. Since we are interested in how
the lift to five dimensions affects the curvature singularities of 
four-dimensional Nernst branes, we work out explicit expressions for
the five- and four-dimensional curvature in our preferred coordinate 
systems. Part of these results have been
obtained in the previous literature, and where results can be compared,
we find agreement. Our own contribution is to 
explicitly demonstrate how singular 
four-dimensional asymptotic hvLif metrics,
when combined with the four-dimensional scalars encoding
the dynamics of an additional compact direction, lift consistently to
a geometry asymptotic to 
AdS$_5$. This does not only show that the hvLif singularities are
artefacts resulting from, as one might say,
a `bad slicing' of AdS$_5$, but the mechanism is dynamical in the
sense that the run-away of the four-dimensional scalars encodes
the decompactification of the fifth dimension.

This removes all the sp curvature singularities and 
run-away behaviour of scalars that four-dimensional
Nernst branes exhibit at asymptotic infinity.\footnote{Following the 
terminology of \cite{Hawking:1973uf} sp singularities correspond to
a scalar invariant formed out of the Riemann tensor becoming infinite,
while pp singularities are curvature singularities observed in a 
parallely propagated frame. These can occur even if all scalar
curvature invariants are finite, and correspond to infinite tidal forces
experienced by freely falling observers. This will be demonstrated in
some detail later in the paper.}  
In addition,
four-dimensional Nernst branes also have pp curvature singularities
which lead to infinite tidal forces acting on freely falling
observers. These occur at the horizon, and only at zero temperature.
They are again accompanied by run-away behaviour of the scalar fields,
which encodes the
dynamical decompactification of a fifth dimension. But in contrast to
what happens at asymptotic infinity under decompactification, 
the pp singularity of the asymptotic hvLif space \cite{Copsey:2012gw} 
is not removed but 
lifted to the pp singularity 
of a five-dimensional Kaigorodov-type space-time \cite{Podolsky:1997ik}.
This shows that pp singularities and infinite tidal forces are
intricately related to the vanishing of the entropy (density), thus 
bringing us back full
circle to the third law. We will continue this discussion 
in the final section.

\subsection*{Outline of the paper}

In Section 2 we review 
five-dimensional
$\mathcal{N}=2$ FI gauged supergravity with vector multiplets
and its dimensional reduction to three Euclidean dimensions.
In Section 3 we obtain five-dimensional Nernst branes by 
solving the three-dimensional effective equations of motion
and lifting the solution back to five dimensions. We solve
the full second order equations of motion but observe that
imposing regularity conditions reduces the number of 
parameters by one half, so that the solution will satisfy 
a unique set of first order equations, despite being non-extremal.
By a coordinate transformation the solution can be brought
to the form of a boosted AdS Schwarzschild black brane, 
and further to a metric of Carter-Novotn\'y-Horsk\'y type.
We work out the thermodynamics in full detail, investigate
the extremal limit, compare geometrical properties to those
of rotating black holes, and analyse the behaviour of curvature.
In Section 4 we perform a reduction to four dimensions and 
show that we recover the four-dimensional Nernst branes 
of \cite{Dempster:2015xqa}. The relations between the geometrical
and thermodynamical properties of five-dimensional and four-dimensional
Nernst branes is worked out in detail. 
In Section 5 we interpret the results, obtain a consistent
picture which ties together five- and four-dimensional Nernst branes,
and discuss its interpretation in the context of the gauge/gravity
correspondence. We also come back to the question of a higher-dimensional
string theory embedding, and use the fact that our solutions 
have the same metric as reduced boosted D3-branes to set up
a gauge/gravity dictionary. We briefly explain how  
the method used in this paper to generate solutions 
can be applied to find more general solutions in the future.
Finally we discuss open questions regarding the fate of pp curvature
singularities and of the third law.

Various technical details have been relegated to appendices. 
Appendix A derives certain re-writings of the scalar potential,
which are used in the main text. Appendix B contains the details
of computing thermodynamic quantities using the quasi-local 
energy-momentum tensor. While in Section 3 the Hawking temperature
is obtained from the surface gravity of the Killing horizon, Appendix
C presents an alternative derivation using the Euclidean approach.
This allows to again compare boosted branes to rotating black holes.
Appendices D and E give the details for computing tidal forces 
in five and in four dimensions, respectively. These details have been 
included to give, in combination with the main text,
a full and self-contained account of curvature in five and four 
dimensions. 
In Appendix F we
spell out the details of a `well known' fact about the normalization
of vector potentials, for completeness, and because we are not
aware of an easily digestible and sufficiently 
detailed explanation in the literature.

\section{$\mathcal{N}=2$ gauged supergravity in five dimensions}

\subsection{Lagrangian of five-dimensional $\mathcal{N}=2$ gauged supergravity with vector multiplets}

We start with the five-dimensional Lagrangian for $\mathcal{N}=2$ gauged supergravity coupled to $n$ vector multiplets \cite{Guenaydin1984}.
Our conventions for the ungauged sector follow those of \cite{Cortes:2009cs}, albeit with the opposite sign for the Einstein-Hilbert term:
\begin{eqnarray}\label{5dLagr}
e_5^{-1}\mathcal{L}_5 &=& 
-\frac{1}{2\kappa^2}R_{(5)}-\frac{3}{4\kappa^2}
a_{ij}(h)\partial_{\hat{\mu}} h^i\partial^{\hat{\mu}} h^j
-\frac{1}{4}a_{ij}(h)\mathcal{F}^i_{\hat{\mu}\hat{\nu}}\mathcal{F}^{j|\hat{\mu}\hat{\nu}} \nonumber \\
& &+\frac{\kappa}{6\sqrt{6}} e_5^{-1} c_{ijk} \epsilon^{\hat{\mu}\hat{\nu}\hat{\rho}\hat{\sigma} \hat{\lambda}} \mathcal{F}^i_{\hat{\mu}\hat{\nu}} \mathcal{F}^j_{\hat{\rho}\hat{\sigma}} \mathcal{A}^k_{\hat{\lambda}} +V_5(h),
\end{eqnarray}
with $\kappa^2=8 \pi G_5$. Here $\hat{\mu},\hat{\nu},\ldots$ are five-dimensional Lorentz indices, while $i,j,\ldots=1,\ldots,n+1$ label the five-dimensional gauge fields.
We use a formulation of the theory where the $n$-dimensional scalar manifold
${\cal H}$ 
is parametrised by $n+1$ scalar fields $h^i$ which are subject to real scale
transformations.  This formulation
is natural in the context of the superconformal calculus and will turn out
to be helpful for finding solutions. The construction of the theory of 
five-dimensional vector multiplets coupled to supergravity using the 
superconformal calculus can be found in  
\cite{Bergshoeff:2002qk, Bergshoeff:2004kh}, while the superconformal 
method in general 
is reviewed in \cite{Freedman:2012zz}. We will in addition 
use the formulation of special real geometry developed in 
\cite{2008arXiv0811.1658A,Mohaupt:2009iq,Cortes:2011aj,Mohaupt:2012tu}.

As explained in more detail 
in \cite{Cortes:2011aj,Mohaupt:2012tu}, 
the scalars $h^i$ are special coordinates
on an open domain  $\cU\subset \mathbb{R}^{n+1}$, which is invariant under 
the action of the group $\mathbb{R}^{>0}$ by scale transformations. 
The manifold $U$ is the scalar manifold of an auxiliary
theory of $n+1$ superconformal vector multiplets, from which 
a theory of $n$ vector multiplets coupled to Poincar\'e supergravity
is obtained by gauge fixing. $\cU$ is a so-called 
conic affine special real (CASR) manifold. This means that it carries 
a Hessian metric which transforms with weight 3 under the 
$\mathbbm{R}^{>0}$-action. When choosing special coordinates, which are
affine coordinates with respect to the Hessian structure and transform
with weight 1 under scale transformations, the
Hesse potential is a homogeneous cubic polynomial, $H(h)= c_{ijk}h^i h^j h^k$.
In this way one recovers the original definition of \cite{Guenaydin1984}.
The physical scalar manifold  of the supergravity theory 
can be identified with the hypersurface $\cH\subset \cU$ defined by
\begin{equation}
\label{HS}
H(h)=  c_{ijk}h^i h^j h^k =1 \;.
\end{equation}
Note that the $\mathbb{R}^{>0}$-action is transverse to this
hypersurface, so that we can identify $\cH \simeq \cU/\mathbb{R}^{>0}$.
This is a real version of the superconformal quotients for 
four-dimensional vector multiplets and for hypermultiplets.

The manifold $\cH$ will be referred to as a  
projective special real (PSR) manifold.
In the Lagrangian (\ref{5dLagr}) we use the special coordinates 
$h^i$, but it is understood that the constraint (\ref{HS}) has
been imposed. Within the superconformal calculus this constraint
is the `D-gauge' which gauge fixes the local dilatations of the 
auxiliary superconformal theory in order to obtain the associated
Poincar\'e supergravity theory in its conventional form. In 
(\ref{5dLagr}) this is reflected by the Einstein-Hilbert term having
its dimension-full prefactor $\sim \kappa^{-2}$, rather than being 
multiplied by a conformal compensator to make it scale invariant. 

In (\ref{5dLagr}) we have chosen to express both the scalar and
the vector couplings using the symmetric, positive definite 
tensor field
\begin{equation}
\label{tensor}
a_{ij}(h)=\frac{\partial^2\tilde{H}}{\partial h^i \partial h^j}
=-2\left(\frac{(ch)_{ij}}{chhh}-\frac{3}{2}\frac{(chh)_i (chh)_j}{(chhh)^2}\right), \qquad
\tilde{H}=-\frac{1}{3}\log H \;.
\end{equation}
Here we use a notation which suppresses indices which are summed over:
$chhh := c_{ijk} h^i h^j h^k$, $(chh)_i := c_{ijk} h^j h^k$, etc. 
The tensor $\partial^2 \tilde{H} = a_{ij}dh^i dh^j$ is
a positive definite Hessian metric with 
Hesse potential $\tilde{H}$ on $\cU$. While it is different from 
the conical Hessian metric $g_{\cU} = \partial^2 H$, which 
has Lorentz signature,
with the negative eigendirection along the orbits of the 
$\mathbbm{R}^{>0}$-action, the pullbacks of both metrics to the
hypersurface $\cH$ agree, so that one can use either to obtain
the positive definite metric $g_{\cH}$ which encodes the 
self-couplings of the $n$ physical scalars. The couplings of 
the $n+1$ physical vector fields are given by the restriction 
of the positive definite metric $a_{ij}$ to $\cH$.

The scalar potential $V_5(h)$ in \eqref{5dLagr} results from an
FI-gauging parametrized by $n+1$ gauging parameters $g_i$. Using 
the expressions of \cite{Ceresole:2000jd} and \cite{Cremonini:2008tw}
we find
\begin{equation}\label{ScalPot}
V_5(h) = 2\cdot 6^{-1/3}\left[(chhh)(ch)^{-1|ij} + 3 h^i h^j\right] g_i g_j.
\end{equation}
We have fixed a convenient normalisation of the gauging parameters $g_i$ by 
comparing the dimensional reduction of \eqref{5dLagr} to the 
four-dimensional scalar potential of \cite{Dempster:2015xqa}, evaluated 
for a ``very special prepotential''
\[
F(X)=-\frac{1}{6}\frac{c_{ijk}X^i X^j X^k}{X^0}\;,
\]
that is a prepotential which can arise by reduction from five to four
dimensions.\footnote{Specifically, comparing to eqn. (30) 
of \cite{Cremonini:2008tw}  we have
$M^I=6^{1/3}h^i$, $A^I=2\cdot 6^{-1/6}\cA^i$, and 
$gP_I =\frac{1}{\sqrt{2}}g_i$.}

We remark that while on the physical scalar manifold we have to impose the 
constraint $chhh = 1$, we have kept factors of $chhh$ explicit 
in \eqref{tensor} and \eqref{ScalPot}. This is useful in keeping track
of the scaling weights of fields, and thus checking that expressions are
consistent with the scaling properties of the corresponding gauge-equivalent
superconformal theory. 
We have chosen our conventions such that the scaling 
weights
of the fields used in (\ref{5dLagr}) match with 
\cite{Bergshoeff:2002qk, Bergshoeff:2004kh}. In particular, we take
\[
w(h^i)=-\frac{1}{2}, \quad
w(c_{ijk})=\frac{3}{2}, \quad
w(g^{\hat{\mu}\hat{\nu}})=2, \quad
w(\cA_{\hat{\mu}})=0, \quad
w(\kappa^{-2})=3, \quad
w(g_i)=3 \;.
\]
As a quick check, note
that the Lagrangian \eqref{5dLagr} has scaling  weight 5, so that the resulting 
action has scaling weight 0.
Further, provided that we include the
appropriate factors of $chhh$, the functions
$a_{ij}$ and $V$ are homogeneous
in $h^i$, even in presence 
of the dimension-full factors $\kappa$ and $g_i$, which appear after imposing
D-gauge.\footnote{See \cite{Bergshoeff:2002qk, Bergshoeff:2004kh} and
\cite{Freedman:2012zz} for more details about the superconformal gauge fixing.} Note that throughout the remainder of this paper, we shall set $\kappa^2=8 \pi G_5=1$.

\subsection{Reduction to three dimensions \label{5dto3d}}

We now want to reduce the five-dimensional theory to three (Euclidean) dimensions.
We make the metric ansatz
\begin{equation}\label{5dmetric}
ds_{(5)}^{2}= 6^{-2/3}\sigma^2 \left(dx^{0}+\cA^0_4 dx^4\right)^{2}
- 6^{1/3}\left(\frac{\phi}{\sigma}\right)\left(dx^4\right)^2 + \frac{6^{1/3}}{\sigma\phi} ds^2_{(3)},
\end{equation}
where all fields depend only on the coordinates of the three-dimensional space.
In addition we choose to switch off all of the five-dimensional gauge fields $\cA^i=0$, i.e.\ we look only for uncharged five-dimensional solutions.\footnote{
We remark that 
four-dimensional solutions which will lift to charged 
five-dimensional solutions have been found in \cite{Dempster:2015xqa}.
The detailed analysis of these solutions is left to future work.
} 
The presence of the  Kaluza-Klein one-form $\cA^0=\cA^0_4 dx^4\equiv -\sqrt{2}\z^0 dx^4$ indicates that we are looking for non-static five-dimensional solutions.
Upon compactification of the $x^0$ circle this will give rise to a non-trivial electric charge for the corresponding four-dimensional solution.
Note that whilst the Killing vector $\partial/\partial x^0$ is always space-like in five dimensions, $\partial/\partial x^4$ can be either time-like, space-like, or null, depending on the magnitude of $\cA^0_4$. 
However, after performing the dimensional reduction over $x^0$, the $x^4$ direction will always be time-like in four dimensions, and so we are able to use the same dimensional reduction technique as in \cite{Dempster:2013mva}, i.e.\ we reduce over both a space-like and time-like direction.

%

The resulting three-dimensional action is given by
\begin{equation}\label{3dLagr1}
e_3^{-1} \mathcal{L}_3 = -\frac{1}{2}R_{(3)} -\frac{3}{4}a_{ij}(h)\partial_\mu h^i\partial^\mu h^j 
-\frac{1}{4\phi^2}(\partial\phi)^2 -\frac{3}{4\sigma^2}(\partial\sigma)^2 +\frac{\sigma^3}{12\phi}(\partial \z^0)^2+V_3(h),
\end{equation}
where the three-dimensional scalar potential is given by
\begin{equation}\label{3dscalpot}
V_3(h)=\frac{6^{1/3}}{\sigma\phi} V_5(h) =\frac{2}{\sigma\phi}\left[(chhh)(ch)^{-1|ij}+3 h^i h^j\right]g_i g_j.
\end{equation}
In order to solve the equations of motion resulting from \eqref{3dLagr1} it turns out to be convenient to introduce the variables $u$, $v$ and $y^i$ via
\begin{equation}\label{fieldredef}
\sigma =u^{-\frac{1}{2}}v^{-\frac{1}{2}}, \quad
\phi =u^{\frac{1}{2}}v^{-\frac{3}{2}}, \quad
y^i=vh^i, \quad
\hat{g}_{ij}(y)=-\frac{3}{4v^2}a_{ij}(h),
\end{equation}
so that the three-dimensional Lagrangian \eqref{3dLagr1} becomes
\begin{equation}\label{3dLagr2}
e_3^{-1} \mathcal{L}_3 = -\frac{1}{2}R_{(3)} +\hat{g}_{ij}(y)\partial_\mu y^i\partial^\mu y^j 
-\frac{1}{4u^2}(\partial u)^2   +\frac{1}{12 u^2}(\partial \z^0)^2    +V_3(y) .
\end{equation}
The scalar potential is given in terms of the new fields by
\begin{eqnarray}
V_3(y) &=& 2\left[(cyyy)(cy)^{-1|ij}+ 3 y^i y^j\right]g_i g_j \nonumber \\
&=& 3\left[\hat{g}^{ij}(y)+4y^i y^j\right] g_i g_j \, .
\end{eqnarray}
The explicit steps used in getting to the second line are carried out in Appendix \ref{App:ScalPot}.

We note that the Lagrangian \eqref{3dLagr2} has no explicit dependence on the 
field $v$ appearing in the metric ansatz. This reflects the fact that when
taking the rescaled scalar fields $y^i$ as independent variables,
the field $v$ becomes dependent, and can be recovered from the equation
\[
v^3 =cyyy,
\]
which follows from the hypersurface constraint $chhh=1$. 
In terms of the new fields $u$ and $v$, the five-dimensional metric ansatz \eqref{5dmetric} becomes
\begin{equation}\label{5dmetric2}
ds^2_{(5)}= \frac{6^{-2/3}}{uv}\left(dx^0 -\sqrt{2}\,\z^0 dx^4\right)^2 - 6^{1/3}\frac{u}{v}(dx^4)^2 +6^{1/3}v^2 ds^2_{(3)}.
\end{equation}
The independent three-dimensional variables are: the metric $ds^2_{(3)}$,
the scalars $y^i$ which encode the $n$ independent five-dimensional
scalars together with the Kaluza-Klein scalar $v$, the second 
Kaluza-Klein scalar $u$, and the scalar $\zeta^0$ which is dual to the
Kaluza-Klein vector from the reduction over $x^4$. 
The metric on the scalar submanifold parametrized by the $y^i$,
\begin{equation}\label{hatgeqn}
\hat{g}_{ij}(y)=\frac{3}{2}\left(\frac{(cy)_{ij}}{cyyy}-\frac{3}{2}\frac{(cyy)_i(cyy)_j}{(cyyy)^2}\right),
\end{equation}
is, up to a constant factor, isometric to the positive definite 
Hessian metric \eqref{tensor} on the manifold $\cU \simeq
\cH \times \mathbbm{R}^{>0} \simeq \cH \times \mathbbm{R}$. 
As shown in \cite{Cortes:2011aj}
this metric is isometric to the product metric $g_{\cH} + dr^2$  
on $\cH \times \mathbbm{R}$. 
From \eqref{3dLagr2} it is manifest
that the scalar manifold $Q$ of the three-dimensional Lagrangian
carries a product metric, with the first factor parametrized by $y^i$ and
the second factor parametrized
by $u$ and $\zeta^0$. By inspection,\footnote{For a systematic 
analysis of the scalar manifolds occuring in reduction to 
three space-like dimensions, we refer the reader to 
\cite{Cortes:2014jha,Cortes:2015wca}.} 
the second factor is 
locally isometric
to the metric of the 
pseudo-Riemannian symmetric space $SU(1,1)/SO(1,1)\simeq AdS_2$,
which can be thought of as the `indefinite signature version' of 
the upper half plane (equivalently, of the unit disk) $SL(2,\mathbbm{R})/SO(2) 
\simeq SU(1,1)/U(1)$. To be precise $u$ and
$\zeta^0$ parametrise an open subset which can identified with the
Iwasawa subgroup of $SU(1,1)$, or, in physical terms, with the static
patch of $AdS_2$. The combined scalar manifold parametrized by 
$y^i, u, \zeta^0$, 
\begin{equation}
\label{Q}
Q = {\cal H} \times \mathbbm{R} \times \frac{SU(1,1)}{SO(1,1)} \;,
\end{equation}
has dimension $n+1+2 = n+3$. 

If we perform the reduction of five-dimensional 
supergravity with $n$ vector multiplets 
to three 
Euclidean dimensions without any truncation, then 
the resulting scalar manifold is a
para-quaternionic K\"ahler manifold $\bar{N}_{PQK}$ of 
dimension $2(2n+2) + 4 = 4n + 8$
\cite{Cortes:2014jha,Cortes:2015wca}. The submanifold $Q$
is obtained by a consistent truncation and therefore it is a totally
geodesic submanifold of $\bar{N}_{PQK}$. 
We remark that $Q$ is a (totally geodesic) submanifold
of the $(2n+4)$-dimensional 
totally geodesic para-K\"ahler manifold
$S_{PK}$ described in \cite{Dempster:2013mva,Errington:2014bta}, 
\[
Q = {\cal H} \times \mathbbm{R} \times \frac{SU(1,1)}{SO(1,1)} \subset
S_{PK} \subset \bar{N}_{PQK} \;.
\]
It was shown there how to obtain explicit stationary non-extremal
solutions  of four- and five-dimensional
ungauged supergravity  by dimensional reduction over time. 
As we will see in the following, it is still possible to obtain 
explicit solutions in the gauged case, where the field equations of the
three-dimensional scalars are modified by a scalar potential. 
While we will retrict ourselves to the submanifold $Q$ in this
paper, the higher dimensional para-K\"ahler submanifold $S_{PK}$ will 
be relevant when the present work is extended to more general, charged
solutions, including the solutions found in \cite{Dempster:2015xqa}.


\section{Five-dimensional Nernst branes}

\subsection{Solving the equations of motion}\label{SSec:Eoms}

We now turn to the three-dimensional equations of motion coming from \eqref{3dLagr2}.
The equations of motion for $y^i,u$ and $\z^0$ read:
\begin{equation}
 \n^2 y^i +\hat{\Gamma}^i_{jk}(y)\partial_\mu y^j\partial^\mu y^k  +3\hat{\Gamma}^i_{jk}(y) \hat{g}^{jm}(y) \hat{g}^{kn}(y) g_m g_n -12 (y^jg_j)\hat{g}^{ik}(y) g_k =0, \label{yeom}
\end{equation}
\begin{equation}
 \nabla^2 u -\frac{1}{u}(\partial u)^2-\frac{1}{3u}(\partial \z^0)^2 =0, \label{ueom}
\end{equation}
\begin{equation}
 \nabla^2 \z^0 -\frac{2}{u}\partial_\mu u\, \partial^\mu \z^0=0, \label{z0eom}
\end{equation}
where we have introduced the Christoffel symbols for the metric $\hat{g}_{ij}(y)$:
\[
\hat{\Gamma}^i_{jk}(y) = \frac{1}{2}\hat{g}^{il}(y) \partial_l\hat{g}_{jk}(y).
\]
Meanwhile, the Einstein equations read
\begin{align}\label{Einseqn}
& -\frac{1}{2}R_{(3)|\mu\nu} +\hat{g}_{ij}(y)\partial_\mu y^i \partial_\nu y^j -\frac{1}{4u^2}\partial_\mu u\,\partial_\nu u +\frac{1}{12u^2}\partial_\mu \z^0 \partial_\nu \z^0
  +3 g_{\mu\nu} \left[\hat{g}^{ij}(y) +4 y^i y^j\right]g_i g_j =0.
\end{align}

We now proceed to solving the equations of motion 
\eqref{yeom}--\eqref{Einseqn}, and
make the following brane-type ansatz for our three-dimensional line element:
\begin{equation}\label{3dmetric}
ds^2_{(3)}=e^{4\psi}d\tau^2 +e^{2\psi}(dx^2+dy^2),
\end{equation}
where $\psi=\psi(\tau)$ is some function to be determined, and $\tau$ is a 
radial coordinate which parametrizes the direction orthogonal to the
world-volume of the brane.   
This is the same brane-like ansatz for the three-dimensional line element 
as in \cite{Dempster:2015xqa}.
Moreover we will impose  that all of the fields $y^i$, $\z^0$ and $u$ depend only on $\tau$.
This coordinate has been chosen such that it
is an affine curve parameter for the curve $C: \tau \mapsto (y^i(\tau), 
u(\tau), \zeta^0(\tau))$ on the scalar manifold $Q$.

The Ricci tensor has components
\[
R_{\tau\tau}=2\ddot{\psi}-2\dot{\psi}^2, \quad
R_{xx}=R_{yy}=e^{-2\psi}\ddot{\psi},
\]
from which we find that the Einstein equations \eqref{Einseqn} become
\begin{equation}\label{xxEE}
V_3(y)=\frac{1}{2}e^{-4\psi}\ddot{\psi},
\end{equation}
for $\mu=\nu\neq \tau$, and
\begin{equation}\label{HC1}
-\frac{1}{2}\ddot{\psi}+\dot{\psi}^2 =-\hat{g}_{ij}(y)\dot{y}^i\dot{y}^j +\frac{\dot{u}^2}{4u^2} -\frac{(\dot{\z}^0)^2}{12u^2},
\end{equation}
for $\mu=\nu=\tau$, where we have used \eqref{xxEE}. 
We will now consider the equations of motion for each of $\z^0$, $u$ and $y^i$ in turn.

\subsubsection*{$\z^0$ equation of motion}

The equation of motion \eqref{z0eom} for $\z^0$ can be brought to the form
\[
\frac{d}{d\tau}\left(\frac{1}{u^2}\dot{\z}^0\right)=0,
\]
which is solved by
\begin{equation}\label{z0soln1}
\dot{\z}^0 =\sqrt{3}Du^2,
\end{equation}
for some integration constant $D$, where we have chosen the factor for later convenience.
Once we solve the equation of motion for $u$ we will further integrate \eqref{z0soln1} to obtain an expression for the Kaluza-Klein vector $\cA^0=-\sqrt{2} \zeta^0$ appearing in the five-dimensional metric.


\subsubsection*{$u$ equation of motion}

Substituting \eqref{z0soln1} in to the equation of motion \eqref{ueom} for $u$ we find
\begin{equation}
\label{u-eq}
\ddot{u}-\frac{1}{u}\dot{u}^2 -D^2 u^3=0.
\end{equation}
Introducing the variable $\chi=u^{-1}$, this becomes
\begin{equation}
\label{w-eq}
\ddot{\chi}-\frac{\dot{\chi}^2-D^2}{\chi}=0 \;.
\end{equation}
By differentiation we obtain the necessary condition
$\dot{\chi} \ddot{\chi} = \chi \dddot{\chi}$, 
which can be integrated to
$\ddot{\chi} = B_0^2 \chi$,
where $B_0$ is a real constant.\footnote{Negative $B_0^2$ would yield
a solution periodic in $\tau$, which we discard.}
 Parametrizing the general solution as
\begin{equation}\label{wsoln1}
\chi(\tau)=A\cosh(B_0\tau) +\frac{B}{B_0}\sinh(B_0\tau),
\end{equation}
with arbitrary constants $A,B$, and substituting back into the
original equation \eqref{w-eq} we find the constraint
\[
D^2=B^2-B_0^2 A^2\;,
\]
which imposes one relation between the four constants $D,A,B,B_0$. 
It will turn out to be useful in what follows to consider $A$, $B_0$ and $\Delta:=B-B_0A$ to be the independent quantities, and to write everything in terms of these.
In particular, we then have $D^2=\Delta(\Delta+2B_0A)$.

We are also now in a position  to further integrate \eqref{z0soln1}, which we write as
\begin{equation}\label{zeta_0_pm}
\dot{\z}^0 =\pm\frac{\sqrt{3\Delta(\Delta+2B_0A)}}{\chi^2}.
\end{equation}
\\
For simplicity we will chose the negative 
sign in~\eqref{zeta_0_pm}, and will not carry through the
corresponding positive solution. Since $\z^0$ is dual to 
a Kaluza-Klein vector, this means that we have fixed the 
sign of the `charge' that the solution carries.\footnote{As 
we will see in the following, the solution carries electric
charge from the four-dimensional point of view and linear
momentum from the five-dimensional point of view.} 
Substituting in \eqref{wsoln1} and integrating, we find 
\begin{equation}
\label{zeta_0_dot}
\z^0 (\tau)= \frac{\sqrt{3}\,B_0\, u(\tau)}{\sqrt{\Delta(\Delta+2B_0A)}}\left[A\sinh(B_0\tau)+\frac{B}{B_0}\cosh(B_0\tau)\right]-\z^0_\infty,
\end{equation}
for some integration constant $\z^0_\infty$, which can be fixed by imposing a suitable physicality condition on the solution.

At this point we anticipate that a horizon, if it exists, will
be located at $\tau \rightarrow \infty$. 
Moreover, as we will show in section \ref{Section:4d}, upon dimensional reduction we obtain a four-dimensional 
stationary (in fact static) solution with a Killing horizon. 
Such horizons admit, for finite temperature, 
an analytic continuation to a bifurcate horizon
\cite{Racz:1995nh}. In order that the four-dimensional one-form  
$\mathcal{A}^0(\tau)$ is well defined, it must vanish at the horizon
\cite{Kobayashi:2006sb,Hartnoll:2009sz}, see also Appendix \ref{Vanishing}.

This fixes
\[
\z^0_\infty = \frac{\sqrt{3}\,B_0}{\sqrt{\Delta(\Delta+2B_0A)}},
\]
and therefore the Kaluza-Klein one-form is given by
\begin{equation}
\cA^0(\tau) = -\sqrt{\frac{6\Delta}{\Delta+2B_0A}}\,u(\tau) e^{-B_0\tau}dx^4.
\end{equation}


\subsubsection*{$y^i$ equation of motion}

The equation of motion \eqref{yeom} for the $y^i$ becomes
\begin{equation}\label{yeom2}
e^{-4\psi}\ddot{y}^i +e^{-4\psi}\hat{\Gamma}^i_{jk}(y) \dot{y}^j\dot{y}^k +3\hat{\Gamma}^i_{jk}(y)\hat{g}^{jm}(y)\hat{g}^{kn}(y)\,g_m g_n -12\,\hat{g}^{ij}(y)\,g_j(y^k g_k)=0.
\end{equation}
To proceed, we first contract \eqref{yeom2} with the dual scalar fields $y_i:=-\hat{g}_{ij}(y)y^j$
and make use of the identity
\[
\hat{\Gamma}^i_{jk}(y)y_i =\frac{1}{2}y_i\hat{g}^{il}(y) \partial_l \hat{g}_{jk}(y)=-\frac{1}{2}y^l\partial_l \hat{g}_{jk}(y)=\hat{g}_{jk}(y),
\]
which follows from the fact that $\hat{g}_{ij}(y)$ is homogeneous of degree $-2$ in the $y^i$. 
We thus find
\[
e^{-4\psi}\ddot{y}^i y_i +e^{-4\psi}\hat{g}_{ij}(y)\dot{y}^i\dot{y}^j +V_3(y)=0,
\]
which upon using \eqref{xxEE} becomes
\begin{equation}\label{cont1}
\ddot{y}^i y_i +\hat{g}_{ij}(y)\dot{y}^i\dot{y}^j =-\frac{1}{2}\ddot{\psi}.
\end{equation}
Given that $\hat{g}_{ij}(y)\dot{y}^j=\dot{y}_i$, we can integrate \eqref{cont1} to find
\begin{equation}\label{cont2}
\dot{y}^i y_i =-\frac{1}{2}\dot{\psi}+\frac{1}{4}a_0,
\end{equation}
for some integration constant $a_0$, where the factor has been chosen for later convenience.
Writing
\[
\dot{y}^i y_i =\frac{3}{4}\frac{(cyy)_i\dot{y}^i}{cyyy} =\frac{1}{4}\frac{d}{d\tau}\left(\log cyyy\right),
\]
we can integrate \eqref{cont2}  further to obtain
\begin{equation}\label{cyyysoln}
\log cyyy =-2\psi +a_0\tau +b_0,
\end{equation}
for an integration constant $b_0$. Again the prefactor has been chosen for later convenience. We now return to the Hamiltonian constraint \eqref{HC1}. Using 
\eqref{wsoln1} and \eqref{z0soln1} this becomes:
\begin{equation}\label{HC2}
-\frac{1}{2}\ddot{\psi} +\dot{\psi}^2 =\frac{1}{4}B_0^2-\hat{g}_{ij}(y)\dot{y}^i\dot{y}^j.
\end{equation}

We then have the following picture.
The solutions $y^i(\tau)$ to \eqref{yeom2} should satisfy the constraints \eqref{HC2} and constraint \eqref{cont2}.
One way to proceed, which is valid for generic five-dimensional models, is to 
set all of the $y^i$ proportional to one another, i.e.\ we put $y^i=\xi^i y$ for some constants $\xi^i$, which satisfy
\[
\hat{g}_{ij}(\xi)\xi^i \xi^j=-\frac{3}{4}.
\]
Note that since the (constrained) scalar fields $h^i$ can be recovered from the $y^i$ via $h^i=(cyyy)^{-1/3}y^i$, we see that this ansatz will result in constant five-dimensional scalar fields.

Using \eqref{cont2} we obtain:
\begin{align}
& \frac{3}{4}\left(\frac{\dot{y}}{y}\right)^2 = -\frac{1}{2}\ddot{\psi}+\dot{\psi}^2 -\frac{1}{4}B_0^2, \label{HC3} \\
& \frac{3}{4}\left(\frac{\dot{y}}{y}\right) = -\frac{1}{2}\dot{\psi}+\frac{1}{4}a_0. \label{cont3}
\end{align}
Eliminating the quantity $(\dot{y}/y)$ from \eqref{HC3}--\eqref{cont3} we obtain an equation for the function $\psi(\tau)$:
\[
\ddot{\psi}-\frac{4}{3}\dot{\psi}^2 -\frac{2}{3}a_0\dot{\psi} +\frac{1}{2}B_0^2 +\frac{1}{6}a_0^2 =0.
\]
This is precisely the same equation as was found in \cite{Dempster:2015xqa}, and so can be solved in the same way by
\begin{equation}
e^{-4\psi}=\alpha^3 e^{a_0\tau} \left(\frac{\sinh(\omega\tau+\omega\beta)}{\omega}\right)^3,
\end{equation}
for some integration constants $\alpha$ and $\beta$, where the quantity $\omega$ is given by
\begin{equation}\label{omegadefn}
\omega^2 :=\frac{2}{3}B_0^2 +\frac{1}{3}a_0^2.
\end{equation}
From this, we can integrate \eqref{cont3} to find
\[
y(\tau)=\Lambda e^{\frac{1}{2}a_0\tau}\left(\frac{\sinh(\omega\tau +\omega\beta)}{\omega}\right)^{\frac{1}{2}},
\]
for some constant $\Lambda$, and hence the $y^i$ are given by
\begin{equation}\label{ysoln1}
y^i(\tau)=\lambda^i e^{\frac{1}{2}a_0\tau}\left(\frac{\sinh(\omega\tau +\omega\beta)}{\omega}\right)^{\frac{1}{2}},
\end{equation}
where we have defined $\lambda^i\equiv \xi^i/\Lambda$.
We finally need to ensure that the solution \eqref{ysoln1} satisfies the original equations of motion \eqref{yeom2}.
This fixes $\lambda^i$ in terms of the gauging parameters $g_i$ and other integration constants as
\begin{equation}\label{lambdaconst}
\lambda^i =\pm  \frac{3\alpha^{3/2}}{8 g_i}.
\end{equation}
Therefore the function $v$ appearing in the line element \eqref{5dmetric2} is given by
\begin{equation}
v(\tau)=(c\lambda\lambda\lambda)^{1/3} e^{\frac{1}{2}a_0\tau}\left(\frac{\sinh(\omega\tau +\omega\beta)}{\omega}\right)^{\frac{1}{2}}.
\end{equation}
The signs in \eqref{lambdaconst} should be chosen such that the function $v(\tau)$ is real and positive for all $\tau > 0$.


At this stage we have six independent integration constants $\alpha,\beta,a_0,A,B_0,\Delta$ which are \emph{a priori} yet to be determined.
However, following \cite{Dempster:2015xqa} we choose to set $\beta=0$ in what follows so that the asymptotic region is at $\tau=0$ and the near horizon region at $\tau\rightarrow\infty$.
We can then scale $\tau$ to set $\alpha=1$.

In order for our solution to make sense as a black brane in five dimensions, we need to impose some physicality constraints.
In particular, we require that the
five-dimensional solution generically
has a finite entropy density.\footnote{Since 
the range of the coordinates $(x,y,x^0)$ is infinite, 
the entropy itself will diverge. By `generic' we mean that we allow
that the solution has a limit, which hopefully
will coincide with the zero temperature limit, where the entropy
density becomes zero.}
Combining the five-dimensional and three-dimensional metric ans\"atze \eqref{5dmetric2} and \eqref{3dmetric} we see that finite entropy density corresponds to a finite value of $v^{3/2}u^{-1/2} e^{2\psi}$ as $\tau\rightarrow\infty$ (i.e.\ at the horizon). 
To leading order we find
\[
\left. v^{3/2}u^{-1/2}e^{2\psi}\right|_{\tau\rightarrow\infty} \sim \exp\left(\frac{1}{4}a_0\tau -\frac{3}{4}\omega\tau +\frac{1}{2}B_0\tau\right).
\]
In order that this be finite and non-zero we therefore require $3\omega=a_0+2B_0$ which, given \eqref{omegadefn}, is equivalent to $a_0=B_0$, further resulting in $\omega=B_0$. Hence, the physicality constraint further reduces the number of independent integration constants by one.

Before moving on to study properties of the solution, we summarise the story so far.
The functions appearing in the five-dimensional line element \eqref{5dmetric2} 
are given by
\begin{align}
v(\tau) &= (c\lambda\lambda\lambda)^{1/3}e^{\frac{1}{2}B_0\tau}\left(\frac{\sinh(B_0\tau)}{B_0}\right)^{\frac{1}{2}}, \label{vsoln1} \\
u(\tau) &= \chi(\tau)^{-1}, \quad \chi(\tau)=A\cosh(B_0\tau)+\frac{B}{B_0}\sinh(B_0\tau), \label{usoln1} \\
e^{-4\psi} &= e^{B_0\tau} \left(\frac{\sinh(B_0\tau)}{B_0}\right)^3, \label{psisoln1} \\
 \cA^0(\tau) &= -\sqrt{\frac{6\Delta}{\Delta+2B_0A}}\,u(\tau) e^{-B_0\tau}dx^4,
\end{align}
whilst
the scalar fields $h^i$ parametrising the CASR manifold are constant and given by
\begin{equation}\label{5dscalars}
h^i=\frac{1}{v}y^i = (c\lambda\lambda\lambda)^{-1/3}\lambda^i
=\frac{1}{g_i}\left(c_{lmn}g_l^{-1}g_m^{-1}g_n^{-1}\right)^{-1/3}
 \, .
\end{equation}

We have therefore found a family of solutions to the equations of motion \eqref{yeom}--\eqref{Einseqn} depending on three non-negative parameters $B_0,\Delta,A$. Since the field equations for the three-dimensional scalars
$y^i(\tau), v(\tau), u(\tau)$ are of second order, and our ansatz
amounts to three independent scalar fields (since the $y^i$ have been
taken to be proportional), we should a priori have expected six independent
integration constants. However, as we have seen, physical regularity 
conditions imposed on the lifted, higher-dimensional solution reduces
the number of integration constants by one half. This is consistent with
physical solutions being uniquely characterised by a system of first order 
flow equations, despite that the equations of motion are of second order, 
as has been observed for other types of solutions before
\cite{Mohaupt:2010fk,Mohaupt:2012tu,Dempster:2013mva,Errington:2014bta}.

We further remark that since the physical five-dimensional scalar fields have
turned out to be constant, their only contribution is to generate an
effective cosmological constant, whose value is determined by the value
of the scalar potential at the corresponding stationary point. Since no
five-dimensional gauge fields have been turned on, our solution, which is
valid for any five-dimensional vector multiplet theory, can therefore
be obtained from an effective action, which only contains the Einstein-Hilbert
term together with a cosmological constant, while the gauge fields and
scalar fields have been integrated out.

\subsubsection*{A coordinate change}

We introduce a new `radial' (more accurately: transversal) 
coordinate $\rho$ via
\begin{equation}
e^{-2B_0\tau} =1-\frac{2B_0}{\rho}\equiv W(\rho),
\end{equation}
so that the near horizon region is at $\rho=2B_0$, and the asymptotic region is at $\rho\rightarrow\infty$.
Hence we can use $\rho$ to analytically continue the solution to the region $0\leq \rho\leq 2B_0$ beyond the horizon.
In terms of $\rho$ we find
\begin{equation}\label{usoln2}
u(\rho)=f(\rho)^{-1}W(\rho)^{1/2}, \qquad f(\rho)=A+\frac{\Delta}{\rho},
\end{equation}
where we have defined $\Delta:=B-B_0 A$. Moreover, we have
\begin{equation}
v(\rho)=(c\lambda\lambda\lambda)^{1/3}(\rho W)^{-1/2}, \qquad
e^{4\psi}=\rho^3 W^2,
\end{equation}
and
\begin{equation}\label{KKVsoln1}
\cA^0(\rho) = -\sqrt{\frac{6\Delta}{\Delta+2B_0A}}\,\frac{W(\rho)}{f(\rho)}dx^4.
\end{equation}

Introducing the notation
\[
\tilde{\lambda}:=\left(\frac{1}{6}c\lambda\lambda\lambda\right)^{1/3},
\]
the five-dimensional line element \eqref{5dmetric2} becomes
\begin{eqnarray}\label{5dmetric2a}
ds^2_{(5)} &=& \frac{\rho^{1/2}}{6\tilde{\lambda}} f(\rho)\left(dx^0 -\sqrt{\frac{6\Delta}{\Delta+2B_0A}}\,\frac{W(\rho)}{f(\rho)}dx^4\right)^2 -\frac{\rho^{1/2} W(\rho)}{\tilde{\lambda} f(\rho)}(dx^4)^2 \nonumber \\
&&  +\frac{6\tilde{\lambda}^2 d \rho^2}{\rho^2 W(\rho)}
+  6\tilde{\lambda}^2 \rho^{1/2}(dx^2+dy^2).
\end{eqnarray}

\subsection{Properties of the solution}\label{SSec:Properties}

Let us now turn to an investigation of the properties of the solutions constructed in Section \ref{SSec:Eoms}, which we recall depend on three independent parameters: $A$, $B_0$ and $\Delta$.
It is instructive to look at the cases $A> 0$ and $A=0$ separately.
Moreover, we focus first on the situation $B_0 >0$, and will comment on the $B_0=0$ case later.

\subsubsection*{Solutions with $B_0>0$ and $A>0$}

In this situation it is convenient to introduce the notation:
\begin{equation}
\tilde{\Delta}:=\frac{\Delta}{2B_0 A}.
\end{equation}
After a suitable scaling of the boundary coordinates, and introducing the new radial coordinate 
$r:=\rho^{1/4}$, 
we can bring the five-dimensional line element \eqref{5dmetric2a} to the form
\begin{eqnarray}\label{5dmetric3}
ds^2_{(5)} &=& \frac{r^2}{l^2} f(r)\left(dx^0 -\sqrt{\frac{\tilde{\Delta}}{1+\tilde{\Delta}}}\,\frac{W(r)}{f(r)}dx^4\right)^2 -\frac{r^2 W(r)}{l^2 f(r)}(dx^4)^2 \nonumber \\
&&  +\frac{l^2 d r^2}{W(r) r^2}
+  \frac{r^2}{l^2}(dx^2+dy^2).
\end{eqnarray}
Here $l$ is defined by
\[
l^2 :=96\tilde{\lambda}^2,
\]
and, as we will see below, is the radius of an asymptotic
$\mathrm{AdS}_5$ space, whilst
\[
W(r)=1-\frac{r_+^4}{r^4}, 
\qquad f(r)=A+\frac{\Delta}{r^4},
\qquad r_+^4 :=2B_0.
\]

In order to interpret our solution, as well as to read off the various thermodynamic quantities associated with it,  it is useful to introduce coordinates in terms of which the line element \eqref{5dmetric3} becomes manifestly asymptotically $\mathrm{AdS}_5$.
We observe that the solution is invariant under the combined
rescalings
\begin{equation}
\label{scaling}
A \rightarrow \lambda A \;,\;\;\;\Delta \rightarrow \lambda \Delta
\;,\;\;\;x^0 \rightarrow \frac{x^0}{\sqrt{\lambda}} \;,\;\;\;
x^4 \rightarrow \sqrt{\lambda} x^4 \;\;\;\;\mbox{where}\;\;\;\lambda>0
\end{equation}
of parameters, with $B_0$ invariant. Note that $\tilde{\Delta}$ is
invariant, so that for $A>0$ we obtain a two-parameter family
of solutions parametrized by $B_0$ and $\tilde{\Delta}$.  
The coordinate transformation
\begin{equation}\label{A>0coordtransf}
t=\frac{1}{\sqrt{A}} x^4, \qquad z=\sqrt{A} x^0-\sqrt{\frac{\tilde{\Delta}}{A(1+\tilde{\Delta})}} \, x^4,
\end{equation}
absorbs $A$ and  brings the metric \eqref{5dmetric3} to the form of a boosted AdS Schwarzschild black brane:
\begin{equation}\label{5dmetric4}
ds^2_{(5)} = \frac{l^2 dr^2}{r^2 W} + \frac{r^2}{l^2} \left[-W\left(u_t\,dt +u_z\,dz\right)^2 
+\left(u_z\,dt + u_t\,dz\right)^2 
+dx^2+dy^2\right] .
\end{equation}
The constants
\begin{equation}
u_t =\sqrt{1+\tilde{\Delta}}, \quad
u_z =\sqrt{\tilde{\Delta}},
\end{equation}
satisfy $u_t^2 -u_z^2 =1$ and parametrise a boost along the $z$-direction.
By taking $r\rightarrow\infty$ one sees that \eqref{5dmetric4} indeed asymptotes to $\mathrm{AdS}_5$ with radius $l$. The constant $\tilde{\Delta}$
parametrizes the boost of the brane, while $B_0$, as we will show below,
is a non-extremality parameter and therefore related to temperature.

This metric can be rewritten by making the following co-ordinate
transformation:
\begin{eqnarray}
r&=& e^{l^{-1} \rho}, \qquad x=l y^1, \qquad y=l y^2
\nonumber \\
t&=& {l \over r_+^2}(u_t-u_z) X -l r_+^2 u_z T, \qquad
z= {l \over r_+^2}(u_t-u_z) X +l r_+^2 u_t T \;,
\end{eqnarray}
to obtain
\begin{eqnarray}
ds_{(5)}^2 &=& e^{-2l^{-1} \rho} dX^2 + e^{2 \ell^{-1} \rho}
\bigg(2 dX dT + r_+^4 dT^2 + (dy^1)^2+(dy^2)^2 \bigg)
\nonumber \\
&+&(1-r_+^4 e^{-4 l^{-1} \rho})^{-1} d \rho^2 \;.
\end{eqnarray}
This metric is the 5-dimensional generalized Carter-Novotn\'y-Horsk\'y metric constructed in \cite{Cvetic:1998jf}.

We further remark that the line element (\ref{5dmetric4}) can be 
further simplified by setting $R=r_+r$, $\tilde{T}=t/r_+$, $X=x/r_+$, 
$Y=y/r_+$, $Z=z/r_+$.
This rescaling corresponds to formally setting $r_+=1$ in the function
$W$ in (\ref{5dmetric4}), thus fixing the coordinate of the horizon 
to $r=1$. 
However, this reparametrization obscures the fact that $r_+$ in 
(\ref{5dmetric4}) encodes the temperature, which, as we will show
later, is defined in a reparametrization invariant way.


\subsubsection*{Solutions with $B_0>0$ and $A=0$}


Let us now look at the case where we take $A=0$, so that $f(\rho)=\Delta/\rho$ in \eqref{5dmetric2a}. In this case, after suitably rescaling the boundary coordinates and introducing the radial coordinate $r$ as before, we find that the five-dimensional line element \eqref{5dmetric2a} becomes
\begin{equation}\label{5dmetricA0}
ds^2_{(5)} = \frac{\Delta}{l^2 r^2}\left(dx^0 -\frac{r^4 W(r)}{\Delta} dx^4\right)^2 -\frac{r^6 W(r)}{\Delta l^2}(dx^4)^2 
 +\frac{l^2 dr^2}{r^2 W(r)} +\frac{r^2}{l^2}(dx^2+dy^2). 
\end{equation}

Making the coordinate redefinition
\[
x^4=\frac{1}{2} (t-z), \qquad
x^0+\frac{r_+^4}{2\Delta}x^4 =t+z,
\]
we can bring the metric \eqref{5dmetricA0} to the form \eqref{5dmetric4} of a boosted AdS Schwarzschild black brane.
The boost parameters are given by
\begin{equation}\label{boostA0}
u_t =\cosh\hat{\beta}, \quad u_z=\sinh\hat{\beta},
\end{equation} 
where the quantity $\hat{\beta}$ is defined via
\[
e^{2\hat{\beta}}=\frac{4\Delta}{r_+^4}.
\]
As in the case $A>0$ we obtain a two-parameter family of black 
brane solutions. For $A=0$ the parameters can be taken to be
$B_0$ (equivalently $r_+$) and $\Delta$. We remark that while both
the cases $A>0$ and $A=0$ can be mapped to two-parameter families
of black branes, both families cannot be related smoothly by taking
$A\rightarrow 0$. 



\subsubsection*{Solutions with $B_0=0$ }

If we take $B_0 \rightarrow 0$ in 
\eqref{5dmetric2a} then the region $0 \leq r \leq (2B_0)^{1/4}$ contracts
to $r=0$, which suggests that 
this limit is the extremal limit. We will show later that $B_0=0$ 
does indeed correspond to vanishing surface gravity, and, hence 
vanishing Hawking temperature, and, moreover, that the solution 
is a BPS solution.
  
For any value (zero or non-zero) of $A$ we can then bring the metric to the form
\begin{equation}\label{ExtMetric}
ds^2_{(5)|\mathrm{Ext}} =\frac{l^2 dr^2}{r^2} +\frac{r^2}{l^2}\left[-dt^2 +dx^2 +dy^2+dz^2 +\frac{\Delta}{r^4}(dt+dz)^2\right].
\end{equation}
This solution agrees with the five-dimensional extremal 
Nernst branes found in \cite{BarischDick:2012gj}.\footnote{However,
the `heated up' branes of \cite{Goldstein:2014qha} appear to be different 
from our non-extremal solutions.
} 
We can equivalently obtain this form of the metric from the boosted black brane \eqref{5dmetric4} by taking the limits 
\begin{equation}
\label{ExtLimit}
r_+\rightarrow 0, \qquad
u_t\rightarrow \infty, \qquad
u_t^2 r_+^4 \rightarrow\Delta=\mathrm{const}.
\end{equation}
In the extremal limit $\Delta$ can be interpreted as a boost parameter.
The vacuum $\mathrm{AdS}_5$ solution is obtained by taking the zero 
boost limit  $\Delta\rightarrow 0$. 
Thus in the extremal limit $\Delta$ determines
the mass, or more precisely the mass per worldvolume or tension of the
brane. 
The precise expressions
for the mass and thermodynamic quantities will be calculated in 
Section \ref{SubSec:Thermodynamics5d}.

The metric (\ref{ExtMetric}) displays an interesting
scaling behaviour in the limit $r\rightarrow 0$. To display
it, we introduce coordinates $x^-, x^+$ by\footnote{For $A=1$
these coordinates agree with $x^0$ and $x^4$ in the extremal
limit. Moreover, the near-horizon limit preserves the symmetry
that allows to set $A=1$.}
\[
t = x^+ \;,\;\;\;z = x^- - x^+ \;.
\]
Then the metric becomes
\[
ds^2_{(5)|\mathrm{Ext}} =\frac{l^2 dr^2}{r^2} +\frac{r^2}{l^2}\left[
\left( 1 + \frac{\Delta}{r^4} \right) (dx^-)^2 - 2 dx^-dx^+
+ dx^2 + dy^2 \right] \;.
\]
Dropping terms which are subleading 
in the `near horizon limit' $r\rightarrow 0$ we obtain
\begin{equation}
\label{NH_limit}
ds^2_{(5)|\mathrm{Ext,NH}} =\frac{l^2 dr^2}{r^2} +\frac{r^2}{l^2}\left[
\frac{\Delta}{r^4}  (dx^-)^2 - 2 dx^-dx^+
+ dx^2 + dy^2 \right] \;.
\end{equation}
This metric is invariant under the scale transformations:
\[
x \mapsto \lambda x \;,\;\;\;
y \mapsto \lambda y \;,\;\;\;
r \mapsto \lambda^{-1} r \;,\;\;\;
x^- \mapsto \lambda^{-1} x^- \;,\;\;
x^+ \mapsto \lambda^{3} x^+\;.
\]
Thus the asymptotic metric shows a scaling invariance similar
to a Lifshitz metric with scaling exponent $z=3$ (and no 
hyperscaling violation, $\theta=0$).\footnote{Lifshitz metrics
will be reviewed in Section \ref{Sect:4dNernst}.} 
The only difference is that
the coordinate $x^-$ has scaling weight $-1$ rather than $+1$. 
This type of generalized scaling behaviour was observed in 
\cite{Singh:2010zs,Narayan:2012hk,Singh:2013iba},
where the metric (\ref{NH_limit}) was obtained by taking a particular
limit of boosted D3-branes. We will come back to this 
in Section \ref{conclusions}, where we discuss the dual field theory
interpretation of our solutions.



\subsubsection*{The boosted black brane}

The boosted black brane has similarities with Kerr-like
black holes, with the linear momentum related to the boost
playing a role analogous to the angular momentum. It is instructive
to work this out in some detail, following the discussion 
of the Kerr solution in \cite{poisson2007relativist}.

Let us first look for the existence of static observers, who
remain at constant $(r,x,y,z)$ and as such have velocities
parallel to the Killing vector field $\partial_t$. Therefore static
observers exist in regions where $\partial_t$ is timelike, and the
limit of staticity is at the value of $r$ where
\begin{align*}
g_{tt} = 0 &\Leftrightarrow - W(r) u_t^2 + u_z^2
= 0 
\\ &\Leftrightarrow r^4=u_t^2r_+^4 \geq r_+^4 \;.
\end{align*}
This ``ergosurface'' is always located outside the event horizon, with the trivial exception of globally static (unboosted) spacetimes for which $u_t=1$ and the two surfaces overlap completely. This is different to the rotating case where ergosurface and event horizon always coincide at the north and south pole. 
\\
Beyond the limit of staticity there still exist stationary observers
which are co-moving (more precisely, but less elegantly `co-translating') 
with the brane. Observers which have
fixed $(r,x,y)$ and a constant velocity in the $z$-direction have
world lines tangent to Killing vector fields
\[
\xi_{(v)} = \partial_t + v\, \partial_z \;,
\]
where the quantity $v=\mbox{const.}$ will be referred to as the velocity. Such
co-moving observers exist in regions where $\xi_{(v)}$ is time-like.
Killing vector fields of the form $\xi_{(v)}$ become null for
values of $r$ where
\[
g_{tt} + 2 v g_{tz} + v^2 g_{zz} = 0 \Rightarrow
v_\pm = - \frac{g_{tz}}{g_{zz}} \pm \sqrt{ \left( \frac{g_{tz}}{g_{zz}} \right)^2
- \frac{g_{tt}}{g_{zz}} } \;.
\]
Thus there is a finite range of velocities $v$, given 
by $v_- \leq v \leq v_+$, which co-moving observers can attain. 
Note that at the limit of staticity, where $g_{tt}=0$, 
we find that $v_+ =0$. Therefore $v$ must be negative once
the limit of staticity has been passed. The limit for co-moving
observers is reached when $v_- = v_+ = :w$, which happens at the
point where
\[
g_{tt} g_{zz} - g_{tz}^2 = 0 \;.
\]
It is straightforward to verify that this happens at the 
same value $r_+$ of $r$ where $W(r_+)=0$. The limiting
velocity $w$ is given by
\begin{equation}\label{velocity}
w = - \left. \frac{g_{tz}}{g_{zz}} \right|_{r=r_+} =  
- \frac{u_z}{u_t} \;,
\end{equation}
and can be interpreted as the boost-velocity of the surface
$r=r_+$. Since $W(r_+)=0$ implies that $g^{rr}(r_+)=0$, 
it follows that on this surface outgoing null congruences 
have zero expansion, see \cite{poisson2007relativist} for
the analogous case of a rotating black hole.  
Consequently $r=r_+$ is an apparent horizon,
and since the solution is stationary, an event horizon. Moreover
this event horizon is a Killing horizon for the vector field
$\xi = \partial_t + w \partial_z = \partial_t - \frac{u_z}{u_t} \partial_z$
and we can interpret $w$ as the boost-velocity of this horizon.
Observe that the limit of staticity and the limit of stationarity are
in general different, and only agree in the unboosted limit $u_z=0$ where we 
recover the AdS Schwarzschild
black brane.

We note that there is frame dragging in our solutions,
since the metric is non-static for $u_z\not=0$. 
Indeed, since the metric coefficients are independent of $t$ and
$z$, the covariant momentum components $p_t$ and $p_z$ are conserved.
But even when setting $p_z=0$, particles have a non-vanishing
contravariant momentum component $p^z = g^{zt} p_t \not=0$ 
in the $z$-direction. 
The boost velocity of the metric varies between the horizon 
and infinity. It can be read off by writing the metric in the
form
\[
ds^2_{(5)} = - N^2(r) dt^2 + M^2(r) (dz - v(r) dt)^2 + \cdots
\]
where the omitted terms involve $dx^2, dy^2$ and $dr^2$. 
An observer at fixed $r,x,y$ is co-moving with the space-time
if their velocity is $dz/dt = v$. Bringing the metric \eqref{5dmetric4}
to the above form one finds
\[
v = - \frac{(1-W) u_t u_z}{u_t^2 -W u_z^2 }
\]
with limits
\[
v \xrightarrow[r \rightarrow r_+]{} - \frac{u_z}{u_t} = w \geq - 1\;,
\]
and
\[
v \xrightarrow[r \rightarrow \infty]{} 0 \;.
\]
It is straightforward to check that for $u_t>1$ the boost speed $|v(r)|$ is
strictly monontonically increasing from $|v_\infty|=0$ at infinity to
$|v_{\rm horizon}| = |w |= u_z/u_t\leq 1$ at the horizon. Thus the boost speed
is bounded by the speed of 
light and can only reach it at the horizon and in the extremal 
limit.
Note that the asymptotic AdS space at infinity is not
co-moving. This is different from Kerr-AdS, where the asymptotic
AdS space is co-rotating, with implications for the black brane
thermodynamics \cite{Hawking:1998kw,Caldarelli:1999xj,Gibbons:2004ai}. 
In particular, we will not need to subtract
a background term, corresponding to the asymptotic AdS space,
from our expressions for the boost-velocity 
in order to have quantities satisfying the first law of 
thermodynamics. We will come back to this later when verifying
the first law.

\subsection{BPS solutions \label{Subsection:BPS}}

In this section, we consider the properties of the extremal solution in further detail. We begin by considering the solution ({\ref{ExtMetric}}). 
On making the co-ordinate transformation
\begin{eqnarray}
r=\Delta^{1 \over 4} e^{l^{-1} R}, \qquad x=l \Delta^{-{1 \over 4}} y^1, \qquad y=l \Delta^{-{1 \over 4}} y^2
\nonumber \\
t= {1 \over 2}l \Delta^{-{1 \over 4}}(X-2T), \qquad
z= {1 \over 2}l \Delta^{-{1 \over 4}}(X+2T) \;,
\end{eqnarray}
the metric ({\ref{ExtMetric}}) becomes
\begin{eqnarray}
\label{kaig1}
ds^2 = e^{-2l^{-1} R} dX^2 + e^{2 l^{-1} R} 
\bigg(2 dX dT + (dy^1)^2+(dy^2)^2 \bigg) + dR^2 \;.
\end{eqnarray}
The metric ({\ref{kaig1}}) is a five-dimensional generalized Kaigorodov 
metric, constructed in \cite{Cvetic:1998jf}, which describes gravitational
waves propagating in AdS${}_5$. The supersymmetry of this solution was
investigated in \cite{Cvetic:1998jf}, where it was shown that this solution
preserves 1/4 of the supersymmetry. Furthermore, after making some appropriate co-ordinate transformations, this solution can be shown to correspond to a class of supersymmetric solutions which appears in the classification of supersymmetric solutions
of minimal five-dimensional gauged supergravity constructed in
\cite{nullsusy}. It is straightforward to show that the
null Killing vector which is obtained as a spinor bilinear is
given by $\partial_t-\partial_z$, in the co-ordinates of ({\ref{ExtMetric}}).

The fact that this Killing vector is null rather than timelike is related
to an interesting feature which distinguishes these BPS solutions
from five-dimensional rotating BPS black holes, namely the existence of
an ergoregion, i.e. a region outside the horizon where it is not
possible for observers to remain static. Note that in the 
BPS limit (\ref{ExtLimit}) the limit of staticity is at
$r=\Delta^{1/4} \geq 0$ which is outside the horizon at $r=0$ (unless
we switch off momentum, $\Delta=0$,  and go to global AdS). Therefore
the ergoregion persists in the BPS limit. 

For stationary BPS black holes
the Killing vector obtained as a spinor bilinear is the standard
static Killing vector field $\partial_t$, which is timelike at 
infinity. For rotating black holes $\partial_t$ is different from the
`horizontal' Killing vector field $\partial_t + \omega \partial_\phi$,
which becomes null on the horizon. An ergoregion exists when 
$\partial_t$ becomes space-like outside the horizon. However 
if $\partial_t$ is a bilinear formed out of Killing spinors,
supersymmetry implies that 
it must be either time-like or null. Hence, rotating BPS black holes cannot
have an ergoregion. Moreover it can be shown that the event horizon
of a rotating BPS black hole
must be non-rotating \cite{Cvetic:2005zi}. As we have shown above,
this is different for the extremal limit of an AdS-Schwarzschild
black brane, which is a BPS wave solution in AdS$_5$: the ergoregion 
persists in the BPS limit, and the (degenerate limit of the)   
horizon\footnote{We will show later that in this limit the metric
develops a singularity at the horizon, corresponding to freely
falling observers experiencing infinite tidal forces.} moves with
the speed of light, since $w=-u_z/u_t \rightarrow -1$.
This is consistent with the solution being BPS,
because the Killing vector obtained as a Killing spinor bilinear 
is not $\partial_t$, which is timelike at asymptotic infinity and
becomes spacelike before the horizon, 
but $\partial_t - \partial_z$, which is null everywhere for the 
BPS solution. Moreover, the horizon turning into a purely left-moving
wave is consistent
with the familiar string theory description of a BPS state as a state 
with massless excitations moving in one direction only.

\subsection{Thermodynamics \label{SubSec:Thermodynamics5d}}

We turn to an investigation of the thermodynamics of the black brane solutions of Section \ref{SSec:Properties}. The
Hawking temperature is related to the surface gravity by 
$T=\frac{\kappa}{2\pi}$, where the surface gravity $\kappa$ of a Killing
horizon is given by
\begin{equation}
\label{kappa_squared}
\kappa^2 = \left. - \frac{1}{2} \nabla^\mu \xi^\nu \nabla_\mu \xi_\nu 
\right|_{r=r_+} \;.
\end{equation}
Evaluating this for $\xi = \partial_t + w 
\partial_z$, we find the Hawking temperature $T$:
\begin{equation}\label{Ex:temperature}
\pi T  = 
\frac{r_+}{l^2\,u_t} \;.
\end{equation}
We remark that the same result can be obtained by imposing that the
Euclidean continuation of the solution does not have a conical 
singularity at the horizon, see Appendix \ref{App:Euclideanisation}.

In the zero boost limit
$u_t=1$, $u_z=0$,  we obtain the Hawking temperature
of an AdS Schwarzschild black brane. 
In the extremal limit (\ref{ExtLimit}), where
the boost  parameters go to infinity $u_t,\,u_z \rightarrow \infty$,
while $r_+\rightarrow 0$, the Hawking temperature
becomes zero, $T \rightarrow 0$, irrespective of whether we
keep $\Delta$ finite or not.

Since our solutions are not asymptotically flat, but rather asymptotic
to AdS$_5$, we cannot apply the standard ADM prescription to compute
the mass and linear momentum of our branes. Instead, we 
use the method based on the quasilocal stress tensor 
\cite{Balasubramanian:1999re}, see also \cite{Ambrosetti:2008mt} for
a review in the context of the fluid-gravity correspondence. 
Here we simply present the result, and relegate explicit calculational details to Appendix \ref{App:FluidGrav}.
To leading order in $1/r$ we find that the quasilocal stress tensor takes the form
\begin{equation}\label{Ex:stresstensor}
T_{\mu\nu}= \frac{r_+^4}{2l^3 r^2}\left(\eta_{\mu\nu} +4u_\mu u_\nu\right) +\ldots, 
\end{equation}
where $\eta_{\mu\nu}$ is the Minkowski metric on $\partial\cM_r$ 
which denotes the hypersurface $r={\rm const}.$ of our space-time, 
with coordinates $(t,x,y,z)$. As indicated we have omitted 
terms subleading in $1/r$, since we are ultimately interested in 
expressions which are finite on the boundary $\partial \cM = 
\lim_{r \rightarrow \infty} \partial \cM_r$ of space-time.  
Note that $T_{\mu \nu}$ takes the form of the stress energy tensor of 
a perfect ultra-relativistic
fluid (equation of state $\rho=3p$, where $\rho$ is the energy density and
$p$ is the pressure), with pressure proportional to $r_+^4\sim T^4$.
The proportionality between $r_+$ and $T$ is the same behaviour as 
for large AdS-Schwarzschild black holes. In the absence of a boost,
it is known that AdS-Schwarzschild black branes behave thermodynamically
like large (rather than small) AdS-Schwarzschild black holes 
\cite{AmmonErdmenger:2015}.

Having obtained the quasilocal stress tensor, mass and linear 
momentum can be computed as conserved charges associated to the Killing
vectors of our solution. 
Again, the details are relegated to the appendix \ref{App:FluidGrav}. 
The mass, which 
is the conserved charge associated with time translation invariance, 
is
\begin{equation}\label{Ex:mass}
M=\frac{(4u_t^2-1)r_+^4}{2 l^5}V_3,
\end{equation}
where $V_3=\int_\Sigma d^3x$ is the spatial volume of the brane,
computed with the standard Euclidean metric $dx^2 + dy^2 + dz^2$. Due 
to the infinite extention of the brane, the mass is infinite, and to 
obtain a finite quantity we must either compactify the world volume
directions or define densities. We will do the latter by consistently
splitting off a factor $V_3$ from all extensive quantities.

Next we calculate the momentum in the $z$-direction, which 
is the conserved charge associated to $z$-translation invariance.
The result is
\begin{equation}\label{Ex:momentum}
P_z=-\frac{4r_+^4 u_t u_z}{2 l^5}V_3,
\end{equation}
and vanishes as expected in the zero boost limit $u_z=0$, $u_t=1$. Notice that these charges satisfy $P_z=M \left( -\frac{4 u_t u_z}{4 u_t^2-1} \right)$, which resembles the motion of a non-relativistic body of mass $M$, moving at velocity $v_z=-\frac{4 u_t u_z}{4 u_t^2-1}$. 

Finally, we calculate the Bekenstein-Hawking 
entropy of the solution by integrating
the pull back of the metric over the horizon. Recalling that we are working in units where $8 \pi G_5=1$, we find
\begin{equation}
\label{S}
S=\frac{1}{4G_5}\int_{\Sigma_{r=r_+}}d^3x\sqrt{\sigma} = 2 \pi \int_{\Sigma_{r=r_+}}d^3x\sqrt{\sigma} =\frac{2 \pi r_+^3}{l^3} u_t 
V_3\;,
\end{equation}
where $\sigma$ denotes the pullback of the metric to the surface $\Sigma_r$. 

Using these, we can check that the thermodynamic variables  satisfy the first law:
\begin{equation}\label{Ex:1stlaw}
\delta M=T\delta S +w\, \delta P_z \;.
\end{equation}
We remark that obtaining \eqref{Ex:1stlaw} is a non-trivial
consistency check for the correctness of the definition of the
thermodynamical quantities, which are initially ambiguous because they  
require background subtractions
corresponding to renormalization of the boundary CFT
\cite{Balasubramanian:1999re}, see also 
\cite{Gibbons:2004ai} for a discussion 
in the context of rotating black holes in higher than four dimensions. 
As noted before we do not need to apply a background subtraction for the 
translation velocity $w$, since the asymptotic AdS$_5$ background is
not co-translating. This is different for AdS-Kerr-type black holes, 
where the subtraction of the background rotation velocity is crucial
for obtaining the correct thermodynamic relations 
\cite{Hawking:1998kw,Caldarelli:1999xj,Gibbons:2004ai}.
We also note that $T,M,P_z,S$ which are all defined in a reparametrization
invariant way, depend on the parameter $r_+$, which is therefore a
physical parameter, despite the fact that it could be absorbed into
the coordinates in the line element (\ref{5dmetric4}). Moreover, 
without the ability of varying this parameter, one could not obtain
the temperature/entropy term in the first law. We refer to appendix
\ref{MMCC}
for further details on this technical point.

The extremal limit of these quantities can be reached by taking $r_+\rightarrow 0$ and $u_t\rightarrow\infty$ with $u_t^2 r_+^4\rightarrow\Delta$ fixed.
In this case we find that the entropy density $s:=S/V_3$ vanishes in 
the extremal limit,
$s\rightarrow 0$ as $T\rightarrow 0$. Therefore our solutions
satisfy the strong version of the Nernst law, and will be referred to 
as Nernst branes.\footnote{Incidentially, this version of the Nernst law
is due to Planck, but clearly `Planck brane' would be a bad choice
of terminology.}
Moreover, since in this case $w=-1$, we find $M=|P_z|$, which is of
course the saturation of the BPS bound, as it must be given the
results of Section \ref{Subsection:BPS}.
As already remarked earlier, in the extremal limit the
boost parameter $\Delta$ controls the mass, and $\Delta \rightarrow 0$
is the limit where the solution becomes globally AdS$_5$.


We can eliminate the quantities $r_+$ and  $u_t$ in favour of the thermodynamical variables $T$ and $w$ via
\[
u_t=\frac{1}{\sqrt{1-w^2}}, \quad
u_z=-\frac{w}{\sqrt{1-w^2}}, \quad
r_+ =\frac{l^2(\pi  T)}{\sqrt{1-w^2}}.
\]
In terms of $T$ and $w$ the mass of the solution is given by
\begin{equation}\label{Ex:massTmu}
M(T,w) =\frac{l^3}{2}V_3\left(\frac{3+w^2}{(1-w^2)^3}\right)(\pi T)^4.
\end{equation}
Hence, we see that the heat capacity
\begin{equation}
C_T \equiv \left. \frac{\partial M}{\partial T}\right|_{w} >0,
\end{equation}
is positive, and  the solution is thermodynamically stable. 
This is as expected, at least in the absence of a boost, since it
is well known that AdS-Schwarzschild black branes behave thermodynamically
like large AdS-Schwarzschild black holes \cite{AmmonErdmenger:2015}. 
As we see from \eqref{Ex:massTmu}, the introduction of a boost does not
introduce thermodynamic instablility.

Expressing the entropy in terms of $(T,w)$ we find
\begin{equation}\label{Ex:entropyTmu}
S(T,w)=2 \pi l^3 V_3\frac{(\pi T)^3}{(1-w^2)^2}.
\end{equation}
Note that turning off the boost $u_z=0$, which corresponds to $w=0$, we have $S\sim T^3$, which is the scaling behaviour expected for an $AdS_5$ Schwarzschild black brane.

Indeed we can use \eqref{Ex:entropyTmu} to investigate the behaviour of $S$ as a function of $T$ in both the high temperature and low temperature limits.
The limit of high temperature (equivalently small boost velocity) is
\[
u_z\rightarrow 0, \qquad r_+\rightarrow \infty, \qquad
u_z^2 r_+^4\rightarrow \Delta=\mathrm{const}.
\]
This corresponds to $|w|\ll 1$, and so we see from \eqref{Ex:entropyTmu} that $S\sim T^3$.
The limit of low temperature (equivalently boost velocity approaching the speed of light) is the extremal limit
\[
u_t\rightarrow \infty, \qquad
r_+\rightarrow 0, \qquad
u_t^2 r_+^4 \rightarrow\Delta=\mathrm{const}.
\]
In this case, one can see that $1-w^2 \sim T^{4/3}$, and so the entropy scales like $S\sim T^{1/3}$. 
This is the behaviour predicted for five-dimensional lifts 
of four-dimensional Nernst branes \cite{Dempster:2015xqa}.
We will comment further on the thermodynamic properties of our
solutions  in Section~\ref{conclusions}.

\subsection{Curvature properties of five-dimensional Nernst branes}\label{SSec:5dsingular}

One motivation of the present work is the singular behaviour of the
four-dimensional Nernst branes found in \cite{Dempster:2015xqa}. We will
show in Section \ref{Section:4d} that the five-dimensional Nernst branes
found above are dimensional lifts of these four-dimensional Nernst branes.
To investigate the effect of dimensional lifting on such singularities, we now examine the behaviour of curvature invariants and tidal forces
of the five-dimensional solutions. 
From both the gravitational point of view,
and with respect to applications to gauge-gravity dualities, one would
like the solutions to have neither naked singularities, nor null 
singularities (singularities coinciding with a horizon), while the 
presence of singularities hidden behind horizons is acceptable. In
practice, the presence of large curvature invariants or large tidal forces
will also be problematic, given that the supergravity action we 
start with needs to be interpreted as an effective action. Therefore
large curvature invariants or tidal forces are indications that this
effective description breaks down due to quantum or, assuming an
embedding into string theory, stringy corrections. This might also
limit the applicability of gauge-gravity dualities to only part of
the solution, where the corrections remain sufficiently small.

\subsubsection*{Curvature invariants}

For our five-dimensional metric \eqref{5dmetric4} we compute the Kretschmann scalar and Ricci scalar to be
\begin{equation}
K= \frac{2 \left( 9 r_+^8 -24 r_+^4 r^4 + 20 r^8 \right)}{r^8l^4}, \qquad R= \frac{4 \left( -5r^4+3r_+^4 \right)}{r^4l^2} .
\end{equation}
Note that these only depend on the temperature $T\sim r_+$ and the 
curvature radius $l$ of the AdS$_5$ ground state.
For the extremal solution ($r_+=0$) both curvature invariants take constant values which agree with those for global $AdS_5$:
\[ 
K_{AdS_5} = \frac{2d(d-1)}{l^4}=\frac{40}{l^4}, \qquad R_{AdS_5} = - \frac{d(d-1)}{l^2}=-\frac{20}{l^2} .
\]
For the non-extremal solution 
the curvature invariants tend to the $\mathrm{AdS}_5$ values asymptotically,
but will blow up as $r \rightarrow 0$. Since this is behind the horizon,
there are no naked or null singularities related to the curvature invariants
of five-dimensional Nernst branes.

\subsubsection*{Tidal Forces}

Even if all scalar curvature invariants are finite, there might still be
curvature singularities related to infinite tidal forces. 
Such curvature singularities can be found 
by computing the components of the Riemann tensor in a 
`parallely-propogated-orthonormal-frame' (PPON) associated with the 
geodesic motion of a freely-falling observer. Following \cite{Hawking:1973uf}
they are called pp singularities, in contrast to sp singularities, where
a scalar curvature invariant becomes singular. 
While such singularities
are often considered milder than those associated to curvature invariants,
they are nevertheless genuine singularities and have drastic physical
effects (`spaghettification') on freely falling observers.

The details of this construction for the five-dimensional extremal solution are relegated to Appendix~\ref{App:5dtidalforces}. We only need to consider the
extremal solution, since non-extremal solutions are manifestly analytic
at the horizon $r_+>0$. 
From Table \ref{riemanntable} in Appendix~\ref{App:5dtidalforces} we observe that the non-zero components of the Riemann tensor in the PPON all have near horizon  behaviour of the form
\begin{equation}
\tilde{R}_{abcd} \sim r^{\alpha} \quad \text{with} \quad \alpha \leq 0 \;,
\end{equation}
with $\alpha < 0$ for all but one independent non-vanishing component.
Hence, as the observer approaches the horizon of the extremal brane ($r \rightarrow 0$) these components will diverge, resulting in infalling observers being subject to infinite tidal forces. This is the same behaviour as observed in four dimensions \cite{Dempster:2015xqa},
and seems to be the price for having zero entropy. It is an interesting 
question whether stringy or other corrections could lift this singularity,
and if so, whether it is possible to 
maintain zero entropy.

\section{Four-dimensional Nernst branes from dimensional reduction}\label{Section:4d}

\subsection{Review of four-dimensional Nernst branes \label{Sect:4dNernst}}

We now want to dimensionally reduce our five-dimensional Nernst branes and compare the resulting four-dimensional spacetimes to those found in previous work~\cite{Dempster:2015xqa}. 
Let us therefore review the relevant features, emphasising the problems
that we want to solve. Four-dimensional Nernst branes depend on three
parameters: the temperature $T$, the chemical potential $\mu$ and 
one electric charge $Q_0$. Due to Dirac quantisation\footnote{The 
four-dimensional theory admits both electric and magnetic charges, though for 
the solution in question only one electric charge has been turned on.}
charge is discrete, and the solution depends on two continuous parameters.
The asymptotic geometries, both at infinity and at the horizon, are 
of hyperscaling violating Lifshitz (hvLif) type:
\begin{equation}
\label{hvLif}
ds^2_{(d+2)} = r^{-2(d-\theta)/d} \left( 
- r^{-2(z-1)} dt^2 + dr^2 + \sum_{i=1}^2 dx_i^2 \right) \;.
\end{equation}
Here $t$ is time, $r$ parametrizes the direction transverse to the
brane, and $x_i$, $i= 1, \ldots, d$ are the directions parallel to 
the brane (with $d=D-2$ in $D$ spacetime dimensions). For $\theta=0$
the line element (\ref{hvLif}) is invariant under rescalings
\[
(t,r,x_i) \mapsto (\lambda^z t, \lambda r, \lambda x_i) \;.
\]
The parameter $z$, which measures deviations from `relativistic
symmetry' (due to time scaling different from space) is called the
Lifshitz exponent. For $\theta\not=0$ the metric is not scale invariant
but still scales uniformly, and $\theta$ is known as the hyperscaling
violating exponent. 

For four-dimensional Nernst branes
the geometry at infinity is independent of the temperature.
For finite chemical potential it 
takes the form of conformally rescaled AdS$_4$, which is of the above
type, with $z=1$, $\theta=-1$. Moreover, the curvature scalar goes to
zero $R_{(4)} \rightarrow 0$, while the scalar fields $z^A, A=1\ldots
n_V^{(4)}$ run off to infinity $z^A\rightarrow \infty$. In contrast,
for infinite chemical potential the geometry at infinity is 
asymptotic to hvLif with  $z=3$ and $\theta =1$. The behaviour of 
curvature and scalars is precisely the opposite as previously:
the curvature scalar diverges $R_{(4)}\rightarrow \infty$, while the 
scalar fields go to zero $z^A\rightarrow 0$. The geometry at the 
horizon is independent of the chemical potential, but depends
on the temperature. For zero temperature the asymptotic geometry
is again hvLif with $z=3$, $\theta=1$, but approaching the `opposite
end' of this geometry, so that the curvature scalar goes to zero
$R_{(4)}\rightarrow 0$. While there is no sp curvature singularity
there remains
a pp curvature singularity at the horizon, that is, 
freely falling observers experience infinite tidal forces.
Simultanously the scalar fields
go to infinity $z^A \rightarrow \infty$. This type of behaviour can
be considered as a generalized form of attractor behaviour
\cite{Goldstein:2009cv}. For finite temperature the geometry takes
the expected form for a non-extremal black brane, the product
of two-dimensional Rindler space with $\mathbb{R}^2$. The scalars
and the curvature take finite values, so that the solutions 
are regular at the horizon for non-zero temperature.

The only element of hvLif holography that we will use is the 
entropy--temperature relation 
\[
S \sim T^{(d-\theta)/z} \;,
\]
valid for field theories with hyperscaling violation \cite{Huijse:2011ef}. 

Since the low-temperature asymptotics of the exact entropy--temperature
relation of four-dimensional Nernst branes is $S\sim T^{1/3}$, which 
matches the scaling properties of the asymptotic zero temperature 
near horzion hvLif geometry with $\theta=1$, $z=3$, gauge/gravity
duality implies the existence of a corresponding three-dimensional
non-relativistic field theory with this scaling behaviour. As the 
solution is charged, the global solution should describe the RG flow
starting from a UV theory corresponding to asymptotic infinity, and
ending with this IR theory. Identifying this UV theory turned 
out to be problematic: the solution at infinity jumps discontinously
between finite and infinite chemical potential, and is singular in 
either case. The more likely candidate (having no curvature singularity)
is the conformally rescaled AdS$_4$, which still does not look like
a ground state, due to the run-away behaviour of the scalars. Moreover,
while the geometric scaling properties indicate an entropy--temperature
relation of the form $S\sim T^3$, the high temperature asymptotic of
the four-dimensional Nernst brane solution is $S\sim T$. As discussed
before, this lead to the conjecture that the solution decompactifies
at infinity, and needs to be understood from a five-dimensional 
perspective.

\subsection{$S^1$ bulk evolution}\label{s1evolutionsection}

\begin{figure}
\includegraphics[scale=0.43,trim={0cm 0cm 0 0cm},clip]{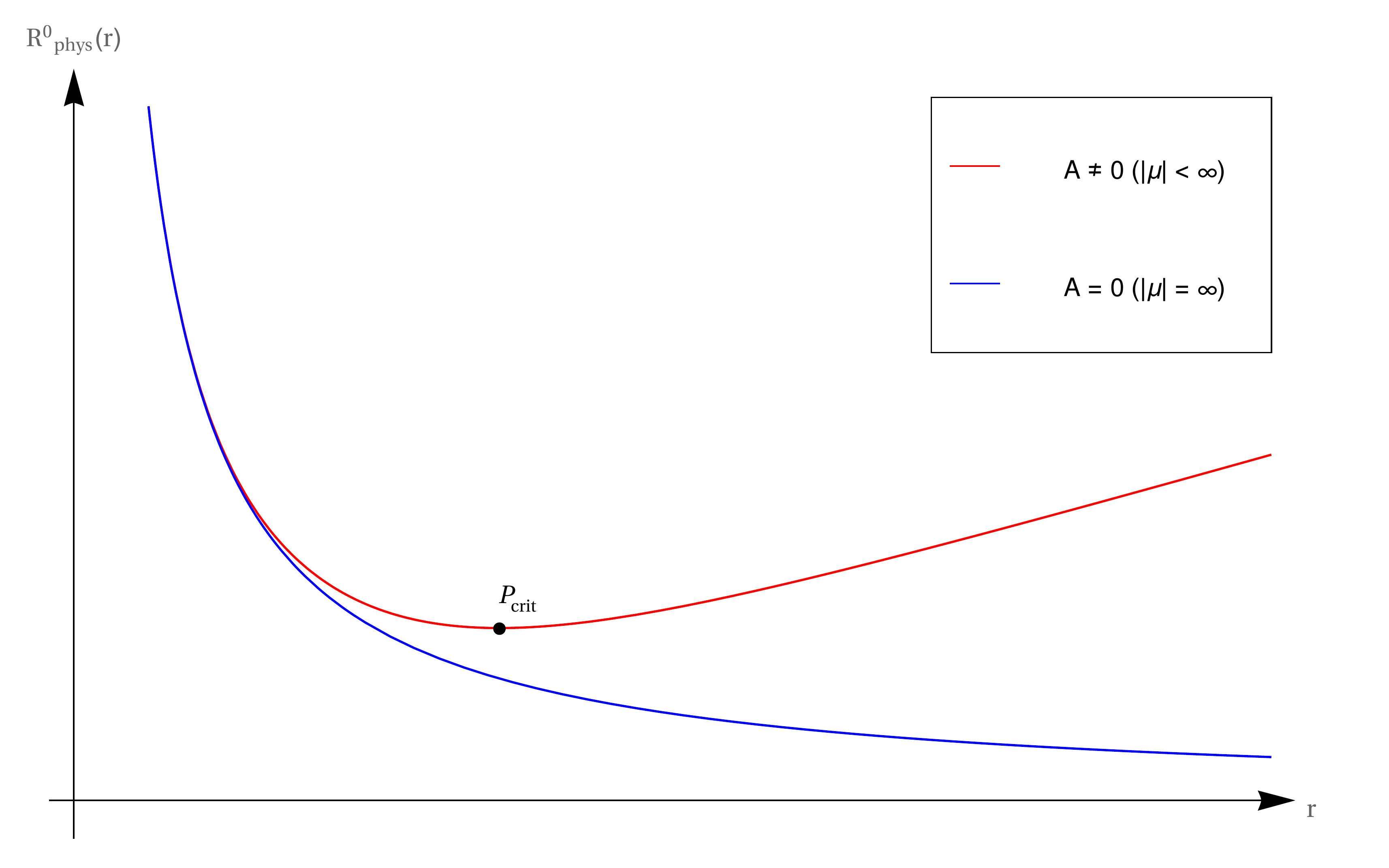}
\caption{Plot showing the evolution of the compactification circle throughout the five-dimensional bulk.}
\label{Fig:S1}
\end{figure}

To relate five-dimensional Nernst branes to four dimensions, we 
make the spacelike direction $x^0$ compact, i.e. we identify $x^0 \sim x^0 + 2 \pi r^0$. Clearly then, to understand the four-dimensional properties, it is crucial to first understand the behaviour of the $x^0$ circle. Writing~\eqref{5dmetric3} as\footnote{As it stands,~\eqref{5dmetric3} is specialized to the case $A>0$ since it involves the variable $\tilde{\Delta}=\frac{\Delta}{2B_0A}$. However, using~\eqref{KKVsoln1}, it is possible to write~\eqref{5dmetric3} in terms of a general Kaluza-Klein vector, valid for both $A>0$ and $A=0$. This then allows the reduction of both cases in parallel, leading to~\eqref{4dmetric}.}
\[
ds^2_{(5)}=e^{2\sigma}(dx^0+\mathcal{A}^0_4 dx^4)^2 +e^{-\sigma}ds^2_{(4)},
\]
with
\begin{equation}\label{circlesizesquared}
e^{2\sigma}=\frac{r^2 f(r)}{l^2},
\end{equation}
we find the four-dimensional line element
\begin{equation}\label{4dmetric}
ds^2_{(4)} = \frac{r}{l}\left\lbrace -\frac{r^2 W(r)}{l^2 f(r)^{1/2}} \, dt^2 + 
f(r)^{1/2}\frac{l^2\,dr^2}{r^2 W(r)} +\frac{r^2}{l^2} f(r)^{1/2} (dx^2+dy^2)\right\rbrace,
\end{equation}
after identifying $x^4\equiv t$. From  \eqref{circlesizesquared} we can
read off the behaviour of the physical (geodesic) length 
$R^0_{\mathrm{phys}}$ of the compactification circle:
\begin{equation}
\label{circlesize}
(R^0_{\mathrm{phys}})^2 = (2\pi r^0)^2 e^{2\sigma}(r) = (2\pi r^0)^2
\left( \frac{A r^2}{l^2} + \frac{\Delta}{r^2 l^2} \right)  \;.
\end{equation}
Thus the geodesic length of the compactification circle $S^1$
varies dynamically along the transverse direction, parametrized by
$r$, of the four-dimensional spacetime, 
as shown in Figure~\ref{Fig:S1}. 
Notice from~\eqref{circlesize} that for $A>0$ there are two competing terms, resulting in decompactification both for $r \rightarrow \infty$ and for $r \rightarrow 0$. The latter decompactification is only reached in the extremal limit, since otherwise we encounter the horizon at $r_+>0$. This implies that in
the non-extremal case the near horizon solution will still depend on the
parameter $A$, while in the extremal case the near horizon solution 
becomes independent of $A$. The insensitivity of the extremal 
near horizon solution
to changes of parameters which determine the asymptotic behaviour at 
infinity, in our case $A$, can be viewed as a version of the
black hole attractor mechanism. Making the solution non-extremal results
in the loss of attractor behaviour by making the near horizon solution 
sensitive to the asymptotic properties of the solution at infinity. 
Of course, the four-dimensional scalars run off to infinity instead of
approaching finite fix-point values, but they do so in a particular,
fine-tuned way, which leads to a consistent lifting of the near horizon
geometry five dimensions. 
A remarkable feature of
solutions with $A>0$ is the existence of a 
critical point, $P_{\text{crit}}$, where the compactification circle reaches a minimal size at $r_{\text{crit}}^4=\Delta/A$. In contrast, for $A=0$, this critical point does not exist and so, whilst the circle continues to decompactify as $r \rightarrow 0$ in the extremal case, it now shrinks monotonically with increasing $r$, ultimately becoming a null circle\footnote{The norm-squared of the
tangent vector $\partial_{x^0}$ goes to zero in this limit.}
of zero size for $r \rightarrow \infty$. This fundamentally different behaviour of the $S^1$ means we must treat the dimensional reduction of the $A>0$ and $A=0$ cases separately in what follows. Additionally, we clearly see that $A$ is the parameter responsible for the asymptotic behaviour at infinity from a five-dimensional point of view. 
This resembles the role played by the parameter $h_0$ in the four-dimensional solutions of~\cite{Dempster:2015xqa}: this connection will be made manifest in the following subsections.

In the case $A>0$, the compactification introduces a new continuous parameter, 
the parametric radius $r^0$ of the circle. We now observe that the
identification $x^0\simeq x^0+2\pi r^0$ breaks the scaling symmetry
(\ref{scaling}), which made the parameter $A$ irrelevant for 
five-dimensional (uncompactified) solutions. For 
$A>0$ there is a circle of minimal size at $r^{4}_{\mathrm{crit}}=\Delta/A$,
with geodesic size $R^0_{\mathrm{crit}}$ given by 
\[
(R^0_{\mathrm{crit}})^2 = 8 \pi \frac{r_0^2}{l^2} \sqrt{\Delta A} \;.
\]
The size of this minimal circle depends only on the combination $r_0^2 \sqrt{A}$ and is therefore invariant under any increase in $A$ that is compensated for by a reduction in $r_0$ and vice-versa. 
This ability to trade $r_0$ for $A$ means that $A$ can be used as the physical parameter controlling the minimal circle size, whilst $r_0$ becomes redundant. It is natural to set $r_0 = \sqrt{A}$, as this is precisely what is needed such that the expression for the four-dimensional charge, $Q_0$, calculated later 
in~\eqref{Q0quant}, is independent of the compactification radius, which is natural
for a quantity which was defined in \cite{Dempster:2015xqa} in a purely
four-dimensional context.

In the case $A=0$, there is no such invariant length and we can see this in a number of ways. Firstly, the $A \rightarrow 0$ limit pushes $r^4_{\text{crit}}= \frac{\Delta}{A} \rightarrow \infty$ and so no minimal circle exists. Secondly, with $A=0$, the geodesic size of the compactification circle is found from~\eqref{circlesize} to be $(R^0_{\text{phys}})^2= \frac{(2 \pi r^0)^2 \Delta}{r^2l^2}$ and depends only on $\Delta$; since this is already a parameter of the five-dimensional solution, there is nothing else to be accounted for and no need for additional parameters. One can try to obtain an invariant length from the size of the circle on the horizon, $R^0_{\text{phys}}(r_+)$, which, assuming non-extremality, will at least be finite. However, it is clear from~\eqref{circlesize} that this will be a function of both $\Delta$ and $r_+$, which again are already existing parameters of the five-dimensional $A=0$ solution.

\subsection{Dimensional reduction for $A>0$ }\label{4dA>0dimredsec}
\subsubsection{Four dimensional metrics and gauge fields}\label{4dmetricmatchsec}

In \cite{Dempster:2015xqa} a family of four-dimensional Nernst 
branes was found, which depend on one electric charge $Q_0$ and
two continuous parameters $B^{(4d)}_0$ and $h_0$, which can be expressed
alternatively in terms of temperature $T^{(4d)}$ and chemical potential
$\mu$. It was also observed that the four-dimensional solutions with finite 
chemical potential exhibited a specific
singular behaviour in the asymptotic regime,
which suggested to be interpreted as a decompactification limit.
Given the behaviour of the  compactification circle, the natural
candidate for a lift of four-dimensional Nernst branes with finite
chemical potential is the $A>0$ 
family of five-dimensional Nernst branes.

We begin by comparing the four-dimensional Nernst brane solutions with finite chemical potential ($h_0\neq 0$) as found in \cite{Dempster:2015xqa} to the four-dimensional metric in~\eqref{4dmetric} obtained by dimensionally reducing our five-dimensional solution with $A>0$.
Setting $\rho=r^4$ in (3.30) of \cite{Dempster:2015xqa} gives:
\begin{equation}
\label{4dNernstBrane}
ds_4^2 = - {\cal H}^{-1/2} W^{(4d)} r^3 dt^2 + \frac{16 {\cal H}^{1/2}}{W^{(4d)}} 
 \frac{dr^2}{r} + {\cal H}^{1/2} r^3 (dx^2 + dy^2) \;,
\end{equation}
where $W^{(4d)}=W^{(4d)}(r) = 1 - \frac{2B^{(4d)}_0}{r^4}$ and
\begin{equation}\label{4dNernstBraneH}
{\cal H}(r) = C \left[ \frac{Q_0}{B^{(4d)}_0} \sinh \frac{ B^{(4d)}_0 h_0}{Q_0} 
+ \frac{ Q_0 e^{-B^{(4d)}_0 h_0/Q_0} }{r^4} \right] =: C {\cal H}_0(r) \;.
\end{equation}
Here $Q_0$ parametrizes the four-dimensional electric charge,
the continuous parameter $h_0 \neq 0$ corresponds to a chemical 
potential $\mu$, with $|\mu|<\infty$\footnote{Due to
the specific choices made for certain signs, the chemical potential
will turn out to be negative. This is correlated with a choice of sign
for the electric charge. There is another branch of the solution,
which we don't give explicitly, where these signs are reversed.},
and the continuous parameter $B^{(4d)}_0\geq 0$ 
corresponds the temperature $T^{(4d)}\geq 0$. The constant $C$ is determined by the choice of 
a prepotential and a gauging of the four-dimensional theory. More precisely, it is determined by the cubic coefficients $c_{ijk}$ and gauging parameters $g_i$, but since we are assuming that this solution can be lifted to five-dimensions, these are the same parameters that enter into our five-dimensional theory in~\eqref{5dLagr}. The precise form of $C$ can be read off from the unnumbered equation between (3.30) and (3.31) in~\cite{Dempster:2015xqa}. At this point we anticipate that the functions $W^{(4d)}$ and $W$ in the four- and five-dimensional solutions can be identified, which allows us to drop the superscrips `4d' 
on $B_0$ and $T$. Since we can no longer rescale the coordinate $r$, matching the coefficients of $dr^2$ 
between the metrics~\eqref{4dmetric} and~\eqref{4dNernstBrane} fixes the relation between the functions $f(r)$ and
${\cal H}(r)$ to be
\[
l^2 f = 16^2 {\cal H} = 16^2 C {\cal H}_0  \;.
\]
Then the remaining metric coefficients match if we rescale 
$t,x,y$ by constant factors involving $l$.\footnote{
Alternatively, we could absorb $l$ into $r$, but then by comparing the
functions $W$ we will conclude that the respective 
parameters $B_0$ differ by a factor $l^4$. Given the relation of $B_0$ 
to the position of the event horizon and to temperature, we prefer
not to do this.}
Writing out the functions $f$ and ${\cal H}$ and comparing, we 
obtain:
\begin{eqnarray}
16^2 C \frac{Q_0}{B_0} \sinh \frac{B_0 h_0}{Q_0} & = & l^2 A \;,\label{match1} \\
16^2 C Q_0 e^{-B_0 h_0/Q_0} &=& l^2 \Delta \;. \nonumber
\end{eqnarray}
While the five-dimensional line element is non-static, the four-dimensional
one is static, but as an additional degree of freedom we have a 
Kaluza-Klein gauge field, given by 
\begin{equation}\label{4dKKtogaugefield}
A^0_t(r) =\z^0 = - \frac{1}{\sqrt{2}} \mathcal{A}^0_4 =
\frac{\sqrt{6}}{\sqrt{2}}\left(\frac{u_z}{u_t}\right) \frac{W(r)}{f(r)} 
=-\frac{\sqrt{3} w W(r)}{f(r)} \;.
\end{equation}
Here we use the definitions and conventions of Section \ref{5dto3d}, and
with regard to four-dimensional quantities, we use the conventions of\cite{Cortes:2009cs}, which were also used in \cite{Dempster:2015xqa}.

By matching the expression for $\dot{\zeta}^0$ given by~\eqref{zeta_0_pm} with the $\tau$-derivative of (3.38) of 
\cite{Dempster:2015xqa}, we can identify the Kaluza-Klein vector with the four-dimensional gauge field provided that
\begin{eqnarray}\label{zeta_0_match}
\frac{\sqrt{3 \Delta (\Delta + 2 B_0 A)}}{A^2} &=&- 
\frac{B_0^2}{2 Q_0 \sinh^2 \frac{B_0 h_0}{Q_0}} \;, \label{match2} \\
1 + \frac{\Delta}{B_0 A} &=& \coth \frac{B_0 h_0}{Q_0} \;. \nonumber
\end{eqnarray}
From this we can find
\begin{eqnarray}
Q_0 &=& -\frac{1}{6} \sqrt{ 3 \Delta (\Delta + 2 B_0 A)}\;, \label{4dchargeA>0}  \\
h_0 &=& \frac{Q_0}{B_0} \mbox{arcoth} \left( 1 + \frac{\Delta}{B_0 A}\right) \;,
\label{match3}
\end{eqnarray}
which expresses the four-dimensional parameters $Q_0,h_0$ in terms
of the five-dimensional parameters $A, \Delta, B_0$. Comparing
(\ref{match1}) to (\ref{match2}) we find that these relations are
mutually consistent provided that 
\begin{equation}\label{normalizations}
16^2 C = - 2 \sqrt{3} l^2 \;.
\end{equation}
This equations relates the overall normalizations of metrics
(\ref{4dmetric}) and (\ref{4dNernstBrane}) and of the underlying
vector multiplet actions.

The four-dimensional chemical potential is given by the asymptotic value of the gauge field $A_t$, which is chosen such that $A_t(r_+)=0$, as explained
in Appendix \ref{Vanishing}. Having matched the five-dimensional Kaluza-Klein vector with the four-dimensional gauge field of~\cite{Dempster:2015xqa}, the corresponding expressions for the chemical potential must also match.\footnote{This can be seen explicitly by applying~\eqref{match2} to (3.39) in~\cite{Dempster:2015xqa} and comparing to the asymptotic value of~\eqref{4dKKtogaugefield}.} For reference, we provide the following expression in terms of both four- and five-dimensional parameters,
\begin{equation}\label{mu}
\mu = \frac{1}{2} \frac{B_0}{Q_0} \left[
\coth \frac{B_0 h_0}{Q_0} - 2 \right] = \frac{\Delta}{2 Q_0 A}
= -\frac{\sqrt{3}}{A} \sqrt{ \frac{\Delta}{\Delta + 2 B_0 A}}\;,
\end{equation}
where we used (\ref{match2}). Notice from~\eqref{4dchargeA>0} that $Q_0<0$ which then forces $h_0<0$ by~\eqref{match3}, which is consistent with the remark in~\cite{Dempster:2015xqa} that $\text{sign}(h_0)=\text{sign}(Q_0)$. Moreover
we observe that $Q_0<0$ implies $\mu<0$. This reflects the correlation in 
the signs
of the charge $Q_0$ and of the chemical potential $\mu$. We have, for
concreteness and simplicity, restricted ourselves to solutions where
$\dot{\zeta}^0>0$, which have turned out to correspond to negative
charge and negative chemical potential. Conversely, solutions with
$\dot{\zeta}^0<0$ will carry positive charge and positive chemical 
potential. This is consistent with the fact that in relativistic
thermodynamics the chemical potentials of particles and antiparticles
differ by a minus sign. 

For completeness we note a few further signs which are implied by our
decisision to focus on solutions with $\dot{\zeta}^0<0$ (and, hence,
$\zeta^0>0$). 
From~\eqref{match1} we deduce that the four-dimensional constant $C$ must be negative, $C<0$, which explains the minus sign in~\eqref{normalizations}. Furthermore, it is clear from~\eqref{4dNernstBraneH} that $\mathcal{H}_0(r)<0$ such that the harmonic function $\mathcal{H}(r)>0$, which we need in order that the roots
of $\mathcal{H}$, which appear in our expression for the solution, are real.

\subsubsection{Momentum discretization, charge quantization and parameter counting}\label{quantizationsection}
Since the reduction is carried out over the $x^0$ direction, it is instructive to calculate the Killing charge associated to the Killing vector $\partial_0=\partial/\partial x^0$. For $A>0$,~\eqref{A>0coordtransf} tells us this is related to the Killing vector $\partial_z$ of the five-dimensional spacetime via
\[
\partial_0 = \sqrt{A} \partial_z.
\]
Since the charge associated with $\partial_z$ is the brane momentum~\eqref{Ex:momentum}, the Killing charge corresponds to momentum in the $x^0$ direction, and can be determined as follows
\begin{equation}
P^0 = \sqrt{A} P_z \simeq -\frac{2}{\sqrt{A}} \sqrt{\Delta(\Delta + 2 B_0 A)}\;,
\end{equation}
where we have omitted $V_3$ and $l$ for simplicity. The periodicity 
of the $x^0$ direction implies that momentum takes discrete values,
\begin{equation}\label{P0quant}
P^0\simeq 
\frac{N}{r^0}= \frac{N}{\sqrt{A}}, \qquad N \in \mathbb{Z}^- \cup \{ 0 \}\;,
\end{equation}
where we have taken into account that $P^0\leq 0$.  Rearranging this as
\begin{equation}
N \simeq \sqrt{A} P^0 \simeq N \simeq - 2 \sqrt{\Delta(\Delta + 2 B_0 A)} 
\end{equation}
and comparing to \eqref{4dchargeA>0}, we see explicitly how the quantization
of the internal momentum implies the quantization 
\begin{equation}\label{Q0quant}
Q_0 \simeq \sqrt{A} P^0 \simeq N, \qquad N \in \mathbb{Z}^- \cup \{ 0 \}
\end{equation} 
of the four-dimensional charge. Note that while the spectrum of 
$P^0$ changes with the radius $r^0=\sqrt{A}$ of the compactification 
circle, the four-dimensional electric charge $Q_0$ is independent
of it. As already mentioned before, $P^0$ and $Q_0$ being negative
results from choosing $\zeta^0$ positive, and solutions with positive
$P^0$ and $Q_0$ can be obtained by flipping signs in (\ref{zeta_0_dot}).
Our choice of signs is consistent with the choices made in 
\cite{Dempster:2015xqa}, in particular the same anti-correlation 
between the signs of $A^0_t = \zeta^0$ and $Q_0$ can be observed
in the equation above (3.38) of \cite{Dempster:2015xqa}.

Let us end this discussion by comparing the number of parameters describing the Nernst branes in different dimensions. Five-dimensional Nernst
branes are parametrized by three continuous paramters $(A,B_0,\Delta)$,
but for $A>0$ we have the scaling symmetry (\ref{scaling}), which tells
us that $A$ is redundant, and that we can parametrize solutions by the two independent and continuous parameters $(B_0,\tilde{\Delta})$, which then correspond to temperature and boost
momentum. Upon compactification a new length scale is introduced that breaks the scaling symmetry present in five dimensions. Consequently, the four-dimensional solution picks up an extra parameter; we need to specify the three independent and continuous parameters $(B_0,\Delta,A)$ in order to completely define the metric~\eqref{4dmetric}. In terms of physical parameters, the four-dimensional solution depends on temperature, charge and chemical potential $(T,Q_0, \mu)$. These are all independent but, as we have seen, since the momentum has a component in the direction we 
compactify over, it becomes discrete, which corresponds directly to the discretization of four-dimensional electric charge. As such, the five-dimensional solution involves two independent and continuous thermodynamic parameters whilst the four-dimensional solution has three independent parameters, two of which are continuous and one of which is discrete.

\subsection{Dimensional reduction for $A=0$ }\label{4dA=0dimredsec}
The two parameter family of four-dimensional Nernst branes found in~\cite{Dempster:2015xqa} exhibits discontinuities in the asymptotic behaviour of both the geometry and the scalar fields when taking the limit $h_0 \rightarrow 0$, or equivalently, $|\mu| \rightarrow \infty$. This discontinuity can be accounted for by the discontinuous asymptotic behaviour of the compactification circle in the limit $A \rightarrow 0$ as seen in Figure~\ref{Fig:S1}. 
We should therefore expect that the infinite chemical potential 
four-dimensional solutions of~\cite{Dempster:2015xqa} with $h_0=0$ 
can be recovered from 
the $A=0$ five-dimensional solution with one dimension made compact.

To demonstrate this relationship we take the four-dimensional Nernst brane metric~\eqref{4dNernstBrane} obtained in~\cite{Dempster:2015xqa} and set $h_0=0$ in~\eqref{4dNernstBraneH} which reduces the function $\mathcal{H}(r)$ to
\[\mathcal{H}(r) = \frac{C Q_0}{r^4}.\] 
Substituting this back into~\eqref{4dNernstBrane} gives the following metric 
\begin{equation}\label{4dNernstBraneA=0}
ds_4^2=-C^{1/2} Q_0^{-1/2}W^{(4d)}r^5 dt^2 + \frac{16 C^{1/2}Q_0^{1/2}dr^2}{W^{(4d)}r^3} + C^{1/2}Q_0^{1/2}r(dx^2+dy^2).
\end{equation}
On the other hand, the dimensional reduction of the $A=0$ class of five-dimensional Nernst branes gives
\begin{equation}\label{4dreducedNernstBraneA=0}
ds_4^2=-\frac{r^5 W}{\Delta^{1/2}l^3}dt^2 + \frac{l \Delta^{1/2}}{r^3 W} dr^2 + \frac{r \Delta^{1/2}}{l^3} (dx^2+dy^2),
\end{equation}
where we have used~\eqref{4dmetric} with $A=0$. Again we identify the functions $W^{(4d)}$ and $W$ appearing in the above metrics, which means the parameters $B_0$ and $T$ will be the same in both cases. As before, this prevents rescaling of the coordinate $r$ and then, by comparing $dr^2$ terms in~\eqref{4dNernstBraneA=0} and~\eqref{4dreducedNernstBraneA=0}, we establish the following relationship between four- and five-dimensional quantities
\begin{equation}\label{match4}
16^2 C Q_0 = l^2 \Delta.
\end{equation}
Again, the remaining metric coefficients can be made to match by rescaling $t,x,y$ by constant factors involving $l$. Following the same procedure as in Section~\ref{4dmetricmatchsec}, we match the gauge field and Kaluza-Klein vector by comparing expressions for $\dot{\zeta}^0$. Specifically, we match~\eqref{zeta_0_pm} with the $\tau$-derivative of (3.38) in~\cite{Dempster:2015xqa}. The two are equivalent provided that 
\begin{equation}\label{match5}
Q_0=-\frac{\Delta}{2 \sqrt{3}},
\end{equation}
which expresses the four-dimensional electric charge in terms of the five-dimensional boost parameter $\Delta$. This is a much simpler expression than in the $A>0$ case and we observe that it matches the $A \rightarrow 0$ limit of~\eqref{match3}. Considering the discontinuities we have encountered previously when taking $A \rightarrow 0$ limits, this seems at first surprising but just reflects
that $Q_0$ is a well defined paramater 
for the four-dimensional solutions of~\cite{Dempster:2015xqa}, 
for any choice of $\mu$ and $T$. Having established $Q_0<0$, we see from~\eqref{match3} that $A \rightarrow 0$ corresponds to $h_0 \rightarrow 0^-$, and thus from~\eqref{mu} that $\mu \rightarrow -\infty$. Lastly, we can substitute~\eqref{match5} into~\eqref{match4} to find the relationship between the overall normalizations of the metrics~\eqref{4dNernstBraneA=0} and~\eqref{4dreducedNernstBraneA=0},
\begin{equation}
16^2 C=-2 \sqrt{3} l^2.
\end{equation}
Clearly this requires $C<0$ as before and, in fact, is exactly the same relationship as for the $A>0$ case in~\eqref{normalizations}, which is expected since $C$ and $l$ are only sensitive to the four- and five-dimensional multiplet actions respectively, and these are indpendent of $A$. Again, since we have matched the gauge fields by comparing $\dot{\z}^0$, the chemical potentials must match and this is indeed the case; using the asymptotic value of~\eqref{4dKKtogaugefield} with $A=0$, we find $\mu=-\infty$ which agrees with the negatively charged, $h_0=0$ solutions in~\cite{Dempster:2015xqa}.


The parameter counting becomes simpler in the $A=0$ case. Five-dimensional Nernst branes are parameterized by two independent and continuous parameters $(B_0,\Delta)$, or equivalently temperature and momentum. However, as we have seen in Section~\ref{s1evolutionsection}, no new length scale is introduced by the reduction and consequently, the four-dimensional solution obtained via dimensional reduction also depends on exactly two independent parameters, $(B_0,\Delta)$, which  are sufficient to completely determine~\eqref{4dNernstBraneA=0} since $A=0$ is fixed. Using~\eqref{match5}, these are equivalent to $(T,Q_0)$ with $\mu=-\infty$. The difference between the five-dimensional and four-dimensional parameters is that the $S^1$ causes charge quantization. This means that whilst both $B_0$ and $\Delta$ are continuous in five dimensions, reducing to four dimensions forces one parameter, namely $Q_0 \sim -\Delta$, to become discrete.

One difference between the $A=0$ solution and the $A>0$ solution is 
that for the $A=0$ solution the compactification circle has no critical
value. Therefore we cannot relate the momentum $P^0$ to the electric
charge $Q_0$ using $r_0$ as a reference scale. This is not a problem
since we could relate $Q_0$ to five-dimensional quantities through
(\ref{match5}), and, moreover, we have seen that the relation between 
$Q_0$ and five-dimensional quanities has a well defined limit for $A\rightarrow
0$. A related feature of the $A=0$ solution is that compactification
circle has no minimal size, and contracts to zero for $r\rightarrow \infty$.
That means that there is a region in this solution, where the circle 
has sub-Planckian, or sub-stringy size. While this is problematic for
an interpretation as a four-dimensional solution, the lifted five-dimensional
solution is simply AdS$_5$, and can be decribed consistently 
within five-dimensional supergravity.

\subsection{Curvature properties of four-dimensional Nernst branes}\label{4dsingularsection}

The four-dimensional solutions with $A>0$ and $A=0$, obtained in Sections~\ref{4dA>0dimredsec} and~\ref{4dA=0dimredsec}, exactly match the $h_0 <0$ and $h_0=0$ solutions of~\cite{Dempster:2015xqa} respectively. In~\cite{Dempster:2015xqa} these four-dimensional solutions were observed to be hyperscaling-violating Lifshitz metrics. It is known from~\cite{Copsey:2012gw} that such solutions suffer from various curvature singularities, and we shall now investigate this by computing the singular behaviour of the metrics \eqref{4dmetric} and~\eqref{4dreducedNernstBraneA=0}.

\subsubsection*{Curvature Invariants}
As with the five-dimensional spacetimes in Section \ref{SSec:5dsingular} we can determine the presence of curvature singularities of our four-dimensional solutions by looking at
the Kretschmann scalar and Ricci scalar associated to the metrics~\eqref{4dmetric} and~\eqref{4dreducedNernstBraneA=0}.
Indeed, since 
any singular behaviour in the curvature will already be present for the extremal solutions, we will concentrate only on the case $r_+=0$. 
The curvature invariants are calculated (using Maple) to be
\begin{align}\label{4dcurvinvs}
K_4^{A>0} &= \frac{r^2  \left( 351 A^4 r^{16} + 1476 A^3 r^{12} \Delta + 2586 A^2 r^8 \Delta^2 + 1284 A r^4 \Delta^3 + 959 \Delta^4 \right)}{4L^2 \left( Ar^4 + \Delta \right)^5} \;, \notag
\\ R_4^{A>0} &= -\frac{3 \left( 15 A^2 r^8 + 34 A r^4 \Delta + 15 \Delta^2 \right)}{2 \sqrt{ \frac{Ar^4 + \Delta}{r^4}} \left( Ar^4 + \Delta \right)^2 rL}\;,  \notag
\\ K_4^{A=0} &= \frac{959r^2}{4 \Delta L^2}, \qquad R_4^{A=0} = - \frac{45r}{2 \sqrt{\Delta}L}  \;.
\end{align}

For $A>0$, or equivalently $|\mu|<\infty$, we find that the Ricci scalar behaves as $R\sim r^{-1}$ for large $r$, and $R\sim r$ for $r\rightarrow 0$, whilst the Kretschmann scalar scales as $K\sim r^{-2}$ and $K\sim r^2$ in these respective regions.
Hence, the curvature invariants will remain finite along the solution.
However, for the $A=0$ solution we will still have the same behaviour at $r\rightarrow 0$, but asymptotically we find $R\sim r$ and $K\sim r^2$.
We therefore have a naked curvature singularity as we approach the boundary of the spacetime.

\subsubsection*{Tidal Forces}
%
In order to investigate whether the 
four-dimensional solutions of \cite{Dempster:2015xqa} admit infinite tidal forces in the near-horizon regime
we will follow the analysis of \cite{Copsey:2012gw}, albeit considering a slightly simpler set-up in which the infalling observer is moving only in the radial direction i.e.\ has zero transverse momentum. The technical details of this procedure can be found in Appendix \ref{App:4dtidalforces}.

Our results in Tables~\ref{4dAneq0tidalforces} and~\ref{4dA=0tidalforces} show that, for both $A>0$ and $A=0$, there exist components of the Riemann tensor, as measured in the PPON, that diverge as $r\rightarrow 0$. This indicates that the radially infalling observer will experience infinite tidal forces at the extremal horizon, $r_+=0$. As before, tidal forces will remain finite on non-extremal horizons, $r_+>0$.

\subsection{Curing singularities with decompactification}
A summary of the singular behaviour of our four- and five-dimensional solutions can be found in Tables~\ref{4dtable} and~\ref{5dtable}. Notice that since $B_0$ and $A$ control the near horizon and asymptotic geometries respectively, we can use these to catalogue any singularities. We will now explain how the singularities present in the four-dimensional hyperscaling-violating Lifshitz solutions of Section~\ref{4dsingularsection}, except those related to infinite tidal forces at extremal horizons, can be removed by dimensional lifting to the asymptotically AdS solutions of Section~\ref{SSec:5dsingular}. 
\begin{table}
\begin{center}
\begin{tabular}{|c|c|c|c|c|} 
\hline \multirow{2}{*}{$B_0,h_0$}  & \multicolumn{2}{c|}{Near Horizon} & \multicolumn{2}{c|}{Asymptotic}  \\\cline{2-5}  
& \parbox[t]{2.5cm}{\centering Curvature \\ Singularity}  & \parbox[t]{2.5cm}{\centering $\infty$ \\ Tidal Forces} & \parbox[t]{2.5cm}{\centering Curvature \\ Singularity}  & \parbox[t]{2.5cm}{\centering $\infty$ \\ Tidal Forces} \\ \hline 
$B_0=0,A=0$ & $\times$ & $\checkmark$ & $\checkmark$ & $\times$ \\ \hline
$B_0=0,A>0$ & $\times$ & $\checkmark$ & $\times$ & $\times$ \\ \hline
$B_0>0,A=0$ & $\times$ & $\times$ & $\checkmark$ & $\times$ \\ \hline
$B_0>0,A>0$ & $\times$ & $\times$ & $\times$ & $\times$ \\ \hline
\end{tabular}
\caption{Summary of singular behaviour of four-dimensional Nernst brane.}\label{4dtable}
\end{center}
\end{table}

\begin{table}
\begin{center}
\begin{tabular}{|c|c|c|c|c|} 
\hline \multirow{2}{*}{$B_0,A$}  & \multicolumn{2}{c|}{Near Horizon} & \multicolumn{2}{c|}{Asymptotic}  \\\cline{2-5}  
& \parbox[t]{2.5cm}{\centering Curvature \\ Singularity}  & \parbox[t]{2.5cm}{\centering $\infty$ \\ Tidal Forces} & \parbox[t]{2.5cm}{\centering Curvature \\ Singularity}  & \parbox[t]{2.5cm}{\centering $\infty$ \\ Tidal Forces} \\ \hline 
$B_0=0, A=0$ & $\times$ & $\checkmark$ & $\times$ & $\times$ \\ \hline
$B_0=0, A>0$ & $\times$ & $\checkmark$ & $\times$ & $\times$ \\ \hline
$B_0>0, A=0$ & $\times$ & $\times$ & $\times$ & $\times$ \\ \hline
$B_0>0, A>0$ & $\times$ & $\times$ & $\times$ & $\times$ \\ \hline
\end{tabular}
\caption{Summary of singular behaviour of five-dimensional Nernst brane.}\label{5dtable}
\end{center}
\end{table}

\subsubsection{Curvature invariants}
Dimensional reduction relates the five-dimensional Ricci scalar to its four-dimensional counterpart by\footnote{Similarly, the Kretschmann scalars are 
related by $K_5 \sim e^{2 \sigma} K_4$. The appearence of 
the second power of the dilaton reflects the fact that the Kretschmann 
scalar is
quadratic in the curvature.}
\[R_5 \sim e^\sigma R_4.\]
As can be seen from Table~\ref{4dtable} and Table~\ref{5dtable}, the only situation where we encounter a curvature singularity is the asymptotic regime of the four-dimensional solution with $h_0=0$, or equivalently $A=0$. In this instance we have $R_4 \sim r$ from~\eqref{4dcurvinvs} whilst $e^{\sigma} \sim 1/r$ from~\eqref{circlesizesquared} resulting in $R_5$ being asymptotically constant and exactly equal to the value of global $AdS_5$ as seen in Section~\ref{SSec:5dsingular}. Recalling that the dilaton $e^\sigma$ measures the geodesic length of the $x^0$ circle, we can now account for the presence of an asymptotic curvature singularity in this class of four-dimensional Nernst branes. Specifically, the four-dimensional, $\mu=-\infty$, asymptotic curvature singularity emerges 
from a `bad slicing,' of the parent $AdS_5$ hyperboloid by a circle 
that gets pinched at infinity. It was shown previously, that the independent
four-dimensional scalars are all proportional to each other, see formula
(3.29) in \cite{Dempster:2015xqa}. It was also observed that for infinite
chemical potential, 
these scalars approach zero asymptotically. From the five-dimensional
point of view, the single profile of the four-dimensional scalars 
determines the profile of the Kaluza-Klein scalar. Therefore the
four-dimensional scalars approaching zero corresponds to the
shrinking of the compactification circle.
When combining this with the singular behaviour
of the four-dimensional metric, we obtain $AdS_5$.

In the $A>0$, or equivalently $|\mu| < \infty$, case the four-dimensional solution of~\cite{Dempster:2015xqa} is asymptotically conformal to $AdS_4$, or $CAdS_4$ for short. We see from~\eqref{4dcurvinvs} that the curvature invariants of $CAdS_4$ behave as $R_4 \sim 1/r$ and vanish asymptotically. At the same time, this is compensated by $e^\sigma \sim r$ from~\eqref{circlesizesquared}, meaning the circle now blows up at large $r$ such that $R_5$ remains asymptotically constant and equal to $R_{AdS_5}$. Thus, in this case the asymptotic behaviour of the
four-dimensional metric and scalars is reversed compared to the $A=0$
case, but still leads to the same five-dimensional asymptotic geometry
after lifting.

\subsubsection{Tidal forces}
As can be seen from Tables~\ref{4dtable} and~\ref{5dtable}, tidal forces are asymptotically irrelevant\footnote{See Appendices~\ref{App:5dtidalforces} and~\ref{App:4dtidalforces} for reasons why.} and so we are only concerned with the situation near the horizon. It is clear that infinite tidal forces are present at the horizon of the extremal Nernst brane in four-dimensions, and are not removed by dimensional lifting. 
This seems to be the price for obtaining the strong version of Nernst's law.


\section{Summary, Discussion, and Outlook}\label{conclusions}

\subsection{The five- and four-dimensional perspective, and looking 
for a field theory dual}

Let us summarize and discuss our results.
Starting from FI-gauged five-dimensional
supergravity with an arbitrary number of vector multiplets, we 
have obtained a two-parameter family of Nernst branes, labelled by 
temperature and momentum. These solutions interpolate between 
AdS$_5$ and an event horizon, and have an entropy--temperature
relation interpolating between $S\sim T^3$ at high temperature/low boost
and $S\sim T^{1/3}$ at low temperature/high boost. The relation 
$S\sim T^3$ is consistent with the scaling properties of AdS$_5$. 
Given that we are working within five-dimensional gauged
$N=2$ supergravity, the dual 
UV field theory should be a conformally invariant four-dimensional 
$N=1$ field theory. Since the metric is the same as in the duality
between gauged $N=8$ supergravity and $N=4$ Super Yang Mills, one 
might expect it to be a conformally invariant $N=1$ Super Yang Mills
theory or a deformation thereoff, but without having a higher
dimensional embedding which allows one to understand the role of
the parameters $c_{ijk}$ and $g_i$ of the gauge theory, we can't say
much more.

We have seen how the five-dimensional lift of four-dimensional
Nernst branes removes all the
singularities at asymptotic infinity 
as well as the mismatch between geometrical and 
thermodynamic scaling relations.
To understand the variation of the compactification circle along the 
transverse direction, which from the four-dimensional point of view is
encoded in the scalar fields, is crucial. The apparently singular behaviour
of the four-dimensional geometry is exactly compensated for by the
singular behaviour of the scalars, or, put differently, by the
behaviour of the circle one has to add to obtain asymptotically AdS$_5$.
Moreover, the compactification circle also accounts for the
four-dimensional chemical potential, which has no counterpart in the
un-compactified five-dimensional solution. However, once we decide to 
make the boost direction compact the dynamics forces the circle to expand
at both ends, and the resulting minimum introduces a new parameter
which we can relate to the chemical potential. As proposed
in \cite{Dempster:2015xqa}, we can interpret the apparently singular 
UV behaviour of four-dimensional Nernst branes as a dynamical 
decompactification limit, which tells us that the description as
a four-dimensional system breaks down and has to be replaced
by a five-dimensional one.

The five-dimensional 
solution admits a non-trivial extremal limit, where the boost
parameter is sent to infinity, while the momentum (density) is 
kept fixed. The resulting extremal near horizon geometry
should define a field theory with entropy-temperature relation 
$S\sim T^{1/3}$. In the context of boosted D-branes
and M-branes, the proposed interpretation is a conformal field
theory in the infinite momentum frame, which carries a finite
momentum density \cite{Cvetic:1998jf}. Moreover, it was proposed
in \cite{Maldacena:2008wh,Singh:2010zs,Narayan:2012hk,Singh:2013iba}
that the compactification of the 
direction along the boost corresponds to discrete light cone
quantisation. In this respect it is interesting to look at 
the asymptotic scaling symmetries of the five- and four-dimensional
extremal solutions near the horizon. In five dimensions the
metric looks like a Lifshitz metric with $z=3$ and $\theta=0$, 
except that the direction along the boost has weight $-1$ instead
of $+1$. Upon reduction to four dimensions, the asymptotic geometry, and
if we go to infinite chemical potential even the global geometry,
is a hyperscaling violating Lifshitz geometry with $z=3$ and 
$\theta=1$ \cite{Dempster:2015xqa}. 
That is, by reduction over the boost direction 
one trades the non-trivial scaling of this direction
for an overall scaling of the metric. 
Following \cite{Maldacena:2008wh,Singh:2010zs,Narayan:2012hk,Singh:2013iba}
we propose to associate a four- and a three-dimensional field
theory to the near-horizon five- and four-dimensional geometries,
respectively, with the three-dimensional theory encoding
the zero mode sector of the discrete light cone quantisation 
of the four-dimensional theory. Both theories are non-relativistic
with Lifshitz exponent $z=3$,
and supersymmtric with two supercharges.\footnote{According to the 
analysis of \cite{DallAgata:2010gj}, extremal four-dimensional
Nernst branes are BPS.}
 The four-dimensional theory
is scale invariant and arises by deforming a four-dimensional
relativistic $N=1$ supersymmetric theory by a finite momentum
density, while the three-dimensional theory is scale covariant.

\subsection{The fate of the third law}

From a strictly gravitational point of view, one should still
worry about the pp curvature singularities which persist 
in the extremal limit irrespective of whether we consider
four-dimensional or five-dimensional Nernst branes. 
While sometimes considered to be  
`mild,' they are genuine curvature singularities which make
the solution geodesically incomplete. Moreover, they are not
cured by stringy $\alpha'$-corrections \cite{Copsey:2012gw}, and 
strings probing 
pp singularities get infinitely excited \cite{Horowitz:2011gh}. 
While at finite temperature there is technically no singularity,
near extremality objects falling towards the event horizon will still
experience very large tidal forces \cite{Horowitz:1997uc}.
This behaviour is, if not an
inconsistency, at least a sign that the singularity has 
physical relevance. Moreover, the pp singularity is clearly 
caused by the way the metric  complies with the strong version 
of Nernst's law, namely through  a warp factor which scales any finite piece
of the world volume\footnote{Here `finite' refers to the Euclidean
metric defined by the coordinates $x,y,z$, which we use to 
refer extensive quantities to `unit world volume.'} to zero volume.
It is not obvious at all how pp singularities could be removed while 
keeping the strong version of Nernst's law. For small BPS black holes,
$R^2$-corrections remove null curvature singularities,
by making the area finite \cite{Dabholkar:2004dq}. 
But as these singularities are of the sp
type, it is not clear what this implies for pp singularities. 
One example where a pp singularity is removed
is the D6 brane of type IIA supergravity, using an M-theory embedding 
\cite{Gueven:2005px}. The effect of higher curvature corrections on 
pp type singularities has been investigated in 
\cite{Barisch-Dick:2013xga,Hristov:2016vbm}. One can also 
approach the problem from the field theory side. For example,
in \cite{Brecher:2000pa} they study the infinite momentum frame
CFT dual to a boosted brane and find evidence that the 
CFT resolves the geometric singularity. In our case it would
be interesting to understand the dual four-, or possibly,
the three-dimensional IR field theory, and to investigate whether it
is non-singular, and whether its ground state is unique or 
degenerate. And if the ground state is unique, one would need to
understand whether this means that (i) pp-singularities are acceptable, 
(ii) they are not, but the dual field theory can be used to 
construct a `quantum geometry' of some sort, (iii) or if there is
some kind of breakdown of gauge/gravity duality in the extremal limit.
Points (i) and (iii) are not necessarily mutually exclusive, since
one might invoke the process version of the third law to assure that
the extremal limit cannot be reached by any physical process.

\subsection{Constructing solutions}

This paper is part of a series of papers where explicit,
non-extremal solutions of five- and four-dimensional
ungauged and gauged supergravity have been constructed using 
time-like dimensional reduction in combination with special 
geometry \cite{Mohaupt:2010fk,Dempster:2013mva,Errington:2014bta}.
As explained in Section 2, solutions correspond
to curves on a particular submanifold of the para-quaternionic K\"ahler 
manifold obtained by reduction to three dimensions, which 
satisfy the geodesic equation deformed by a potential.
As part of the solution we have obtained an explicit expression for a stationary
point of the five-dimensional 
scalar potential, corresponding to an AdS$_5$ vacuum,
for an arbitrary number of vector multiplets and general FI-gauging. 
While we initially obtain solutions to the full second order field equations,
with the corresponding number of integration constants, we have seen that 
once we impose regularity of the lifted five-dimensional solution 
at the horizon\footnote{This is done in the
generic situation, that is in particular for finite temperature.} the 
number of intergration constants is reduced by one half, so that
the solution satisfies a unique set of first order equations. 
Such behaviour has been observed before, and been interpreted
as a remnant of the attractor mechanism \cite{Mohaupt:2010fk}.\footnote{A 
related idea seems to be that of `hot attractors' \cite{Goldstein:2014gta}.} 
For our five-dimensional solutions the scalars are constant,
so that the only sense in which we have attractor behaviour is that 
the scalars sit at a stationary point of the scalar potential. 
However, from the four- and three-dimensional perspective 
we have scalar fields which need to exhibit a particular,
fine-tuned, asymptotic behaviour at the horizon in order to
make the five-dimensional solution regular. This is very similar
to attractor behaviour, and the effect of reducing the number of
integration constants by one half is the same. Such universal 
features of scalar dynamics deserve further study. 

In the present paper we have made a very particular choice 
of the ansatz, which was tailored to obtaining the five-dimensional
lift of the four-dimensional Nernst branes of \cite{Dempster:2015xqa}. 
In the future
we will study systematically other choices, which will lead to other 
and more general solutions. Already in \cite{Dempster:2015xqa} a 
four-dimensional 
magnetic solution was found, and we expect that it is possible to obtain
dyonic solutions as well. It would also be interesting to revisit the
issue of embeddings into ten- and eleven-dimensional supergravity.

\subsection*{Acknowledgements}

The work of JG is supported by the STFC grant ST/1004874/1.
The work of TM is partially supported by STFC consolidated grant 
ST/G00062X/1. The work of 
DE was supported by STFC studentship ST/K502145/1
and by the Scottish International Educational Trust. 
TM would like to thank Gabriel Cardoso and Suresh Nampuri for 
extensive and helpful discussions about their work on Nernst branes.

\subsection*{Data Management}
No additional research data beyond the data presented and cited in this
work are needed to validate the research findings in this work.


\appendix

\section{Rewriting the scalar potential}\label{App:ScalPot}

Our goal in this appendix is to obtain a workable expression for the scalar potential $V_3$ appearing in \eqref{3dLagr2}.
Let us concentrate on the term $(cyyy)(cy)^{-1|ij}$.
This is to be interpreted as the matrix inverse to $(cyyy)^{-1}(cy)_{ij}$ in the sense that
\begin{equation}\label{inverse1}
(cyyy)(cy)^{-1|ij}\frac{(cy)_{jk}}{cyyy}=\delta^i_k.
\end{equation}
Now, using the expression~\eqref{hatgeqn} for $\hat{g}_{ij}(y)$:
\[
\hat{g}_{ij}(y)=\frac{3}{2}\left(\frac{(cy)_{ij}}{cyyy}-\frac{3}{2}\frac{(cyy)_i(cyy)_j}{(cyyy)^2}\right),
\]
we have 
\begin{equation}\label{inverse2}
\delta^i_k =(cyyy)(cy)^{-1|ij}\left[\frac{2}{3}\hat{g}_{jk}(y) +\frac{3}{2}\frac{(cyy)_j(cyy)_k}{(cyyy)^2}\right] .
\end{equation}
We now introduce the dual scalars $y_i$ via
\[
\partial_\mu y_i := \hat{g}_{ij}(y)\partial_\mu y^j,
\qquad
y_i =\frac{3}{4}\frac{(cyy)_i}{cyyy} =-\hat{g}_{ij}(y)y^j.
\]
Hence, \eqref{inverse2} becomes
\begin{equation}\label{inverse3}
\delta^i_k =(cyyy)(cy)^{-1|ij} \left[\frac{2}{3}\hat{g}_{jk}(y) +\frac{8}{3}y_j y_k\right].
\end{equation}
In other words, the quantity $(cyyy)(cy)^{-1|ij}$ is just the inverse of the term in square brackets in \eqref{inverse3}. Thankfully, the latter is easily invertible. Indeed, we find
\[
\frac{3}{2}\left[\hat{g}^{ij}(y) +2 y^iy^j\right]\cdot \frac{2}{3}\left[\hat{g}_{jk}(y)+4y_j y_k\right]
= \delta^i_k \,.
\]
Hence we can rewrite
\begin{equation}
(cyyy)(cy)^{-1|ij}= \frac{3}{2}\hat{g}^{ij}(y)+3 y^i y^j,
\end{equation}
so that the scalar potential term in \eqref{3dLagr2} becomes
\begin{equation}\label{3dscalpot2}
V_3 = 3\left[\hat{g}^{ij}(y) +4 y^i y^j\right]g_i g_j.
\end{equation}

\section{Quasi-local computation of conserved charges}\label{App:FluidGrav}
%

We use the form of our five-dimensional line element given in \eqref{5dmetric4}, 
which can be rewritten as
\begin{equation}\label{AppEq:5dmetric}
ds^2=
\frac{l^2 dr^2}{r^2 W} +\frac{r^2}{l^2}
\left(\eta_{\mu\nu}+\frac{r_+^4}{r^4}u_\mu u_\nu\right) dx^\mu dx^\nu ,
\end{equation}
where $u_\mu=(u_t,0,0,u_z)$. Note that $u_\mu u^\mu=-1$ so we can interpret this as a velocity vector.
Following the procedure of \cite{Balasubramanian:1999re} we want to calculate the quasilocal stress tensor $T^{\mu\nu}$ associated with the metric \eqref{AppEq:5dmetric}.

\subsection{The quasilocal stress tensor}

Given a timelike surface $\partial\cM_r$ at constant radial distance $r$
we define the metric $\gamma_{\mu\nu}$ on $\partial\cM_r$  via the ADM-like decomposition
\begin{equation}\label{AppEq:ADMmetric}
ds^2=N^2 dr^2 +\gamma_{\mu\nu}(dx^\mu +N^\mu dr)(dx^\nu +N^\nu dr) .
\end{equation}
We define the extrinsic curvature $\Theta^{\mu\nu}$ via
\begin{equation}\label{AppEq:extrinsiccurv}
\Theta^{\mu\nu}:=-\frac{1}{2}\left(\n^\mu \hat{n}^\nu +\n^\nu \hat{n}^\mu\right),
\end{equation}
where $\hat{n}^\mu$ is the outward-pointing normal vector to the surface $\partial\cM_r$.
For solutions asymptoting to $AdS_5$ the procedure of \cite{Balasubramanian:1999re} tells us that the quasilocal stress tensor is then given by\footnote{We remind the reader that in this paper we work in units where $8 \pi G=1$.}
\begin{equation}\label{AppEq:Stresstensor1}
T_{\mu\nu}= \Theta_{\mu\nu}(\gamma)- \Theta(\gamma)\gamma_{\mu\nu}-\frac{3}{l}\gamma_{\mu\nu}-\frac{l}{2}G_{\mu\nu}(\gamma) ,
\end{equation}
where
$\Theta=\gamma_{\mu\nu}\Theta^{\mu\nu}$ is the trace of the extrinsic curvature, and $G_{\mu\nu}$ is the Einstein tensor for $\gamma_{\mu\nu}$.
%

For the case at hand we see that the metric \eqref{AppEq:5dmetric} decomposes according to \eqref{AppEq:ADMmetric} with
\begin{equation}\label{AppEq:BoundaryMetric}
N^2=\frac{l^2}{r^2 W}, \qquad 
N^\mu =0, \qquad
\gamma_{\mu\nu}(r)= \frac{r^2}{l^2}\left(\eta_{\mu\nu}+\frac{r_+^4}{r^4}u_\mu u_\nu\right).
\end{equation}
The unit normal vector $\hat{n}^\mu$ to a surface of constant $r$ is given by
\[
\hat{n}^\mu =\frac{r}{l}W^{1/2}(r)\delta^{\mu,r},
\]
from which we find the extrinsic curvature
\begin{equation}
\Theta_{\mu\nu}=-\frac{r}{2l} \left(1-\frac{r_+^4}{r^4}\right)^{1/2}\partial_r \gamma_{\mu\nu}
=-\frac{r^2}{l^3}\left(1-\frac{r_+^4}{r^4}\right)^{1/2} \left(\eta_{\mu\nu}-\frac{r_+^4}{r^4}u_\mu u_\nu\right).
\end{equation}
In order to calculate the trace of this we need an expression for the inverse metric $\gamma^{\mu\nu}$, which is given by
\begin{equation}\label{AppEq:invgamma}
\gamma^{\mu\nu} =\frac{l^2}{r^2}\left[\eta^{\mu\nu} -\frac{r_+^4}{r^4}\left(1-\frac{r_+^4}{r^4}\right)^{-1}u^\mu u^\nu\right],
\end{equation}
where $u^\mu=\eta^{\mu\nu}u_\nu$, etc.
This can be used to compute the trace of the extrinsic curvature
\begin{equation}\label{AppEq:trace}
\Theta=\Theta_{\mu\nu}\gamma^{\mu\nu}
=-\frac{2}{l}\left(1-\frac{r_+^4}{r^4}\right)^{1/2}\left[2+\frac{r_+^4}{r^4}\left(1-\frac{r_+^4}{r^4}\right)^{-1}\right] .
\end{equation}
Putting all this together, and noting that $G_{\mu\nu}(\gamma)=0$, we can use \eqref{AppEq:Stresstensor1} to find the resulting gravitational stress-energy tensor induced on the boundary $\partial \mathcal{M}_r$,
\begin{equation}\label{AppEq:Stresstensor2}
T_{\mu\nu}= \frac{r_+^4}{2l^3 r^2}\left(\eta_{\mu\nu} +4u_\mu u_\nu\right) +\ldots,
\end{equation}
where the dots represent terms which are subleading in the limit $r\rightarrow \infty$.

\subsection{Mass, momentum and conserved charges \label{MMCC}}

The quasilocal stress tensor \eqref{AppEq:Stresstensor2} can be used to compute well-defined mass and other conserved charges for the spacetime \eqref{AppEq:5dmetric}.
Let $\Sigma$ be a spacelike hypersurface in $\partial\cM=\lim_{r\rightarrow\infty}\partial\cM_r$ and make the ADM decomposition
\begin{equation}\label{AppEq:ADM2}
\gamma_{\mu\nu}dx^\mu dx^\nu =-N_\Sigma^2 dt^2 
+\sigma_{ab}(dx^a +N^a_\Sigma dt)(dx^b+N^b_\Sigma dt),
\end{equation}
where $\{x^a\}$ are coordinates spanning $\Sigma$, which has metric $\sigma_{ab}$.
Let $U^\mu$ be the timelike unit normal to $\Sigma$.
Then for any isometry of $\gamma_{\mu \nu}$, which we take to be generated by a Killing vector $\xi$, we can define a conserved charge $Q_\xi$ by
\begin{equation}\label{conservedcharge}
Q_\xi =\int_\Sigma d^{d-1}x\, \sqrt{\sigma}\left(U^\mu T_{\mu\nu}\xi^\nu\right).
\end{equation}
In particular, the mass of the solution is given by taking $\xi=\partial_t$, whilst the momentum in the direction $x^a$ is given by taking $\xi=\partial_a$.

For the boosted black brane we can make the ADM decomposition \eqref{AppEq:ADM2} of the metric \eqref{AppEq:BoundaryMetric} with
\begin{align*}
& \sigma_{xx}=\sigma_{yy}=\frac{r^2}{l^2}, \qquad \sigma_{zz}=\frac{r^2}{l^2}\left(1+\frac{r_+^4}{r^4}u_z^2\right), \\
& N_\Sigma^z 
=\frac{r_+^4}{r^4}u_z u_t\left(1+\frac{r_+^4}{r^4}u_z^2\right)^{-1}, \\
& N_\Sigma^2 =\frac{r_+^8}{l^2 r^6}u_z^2 u_t^2 \left(1+\frac{r_+^4}{r^4}u_z^2\right)^{-1} +\frac{r^2}{l^2}\left(1-\frac{r_+^4}{r^4}u_t^2\right).
\end{align*}
The timelike unit normal to $\Sigma$ has components
\begin{align*}
& U^t = - \frac{l}{r}\left(1+\frac{r_+^4}{r^4}u_z^2\right)^{1/2} \left(1-\frac{r_+^4}{r^4}\right)^{-1/2}, \\
& U^z = \frac{l r_+^4}{r^5}u_t u_z\left(1+\frac{r_+^4}{r^4}u_z^2\right)^{-1/2} \left(1-\frac{r_+^4}{r^4}\right)^{-1/2}.
\end{align*}

Using these expressions, as well as the components of the quasilocal stress tensor \eqref{AppEq:Stresstensor2}, we can calculate the mass and linear momentum associated with the boosted black brane \eqref{AppEq:5dmetric}.
Taking $\xi=\partial_t$ and $\xi=\partial_z$ we obtain the expressions \eqref{Ex:mass} and \eqref{Ex:momentum} for the mass and linear momentum respectively.

Finally, let us add some further comments on the fact that $r_+$, and hence
temperature, is a physical parameter despite that it can be absorbed 
by rescaling coordinates in (\ref{5dmetric4}). From (\ref{conservedcharge}),
(\ref{kappa_squared}), (\ref{S}) it is manifest that all quantities 
entering into the first law are geometric quantities (norms of vectors fields,
and integrals of functions over submanifolds using the induced metric)
which are independent of the choice of coordinates. Applying the
coordinate transformation $R=r_+ r, \tilde{T}=t/r_+, X=x/r_+, Y=y/r_+, Z=z/r_+$
to these expressions, it is straightforward to see that the parameter 
$r_+$ is not eleminated, but scaled out as an overall prefactor. 
In particular
\[
\partial_t = r_+ \partial_T \;,\;\;\;
\partial_z =  r_+ \partial_Z \;,
\]
while
\[
V_3 = \int_\Sigma dx dy dz = r_+^3 \int_\Sigma dX dY dZ
\]
so that irrespective of our choice of coordinates $T\sim r_+$, $S\sim
r_+^3$, $M\sim r_+^4$ and $P_z \sim r_+^4$. It is precisely this
$r_+$-dependence of the thermodynamic quantities that gives rise to
the correct temperature/entropy term in the first law.
Put differently, when working
in the rescaled coordinates $(\tilde{T},R, X,Y,Z)$ the parameter $r_+$ is 
hidden in the choice of the vector field $\xi$ and the volume $V_3$.

\section{Euclideanisation of the boosted black brane}\label{App:Euclideanisation}

As is well known from the study of Kerr black holes, obtaining the Hawking temperature by Euclidean methods is much more subtle for non-static spacetimes. For this reason, we find it useful to give an explicit demonstration of how this works in the case of boosted (non-static) black branes. The treatment of the linear case given below will be parallel to the analysis of the Kerr black hole in~\cite{poisson2007relativist}.

A Euclidean continuation of the boosted black brane solution 
\eqref{5dmetric4} can be obtained by setting $t=i\tau$ and $u_z=i\beta$,
and taking $\tau$ and $\beta$ to be real. 
Observe that following
the standard treatment of the Kerr solution,
we do not only continue
time but also the `boost parameter' $w=-u_z/u_t$, which is 
analogous to the angular momentum parameter of the Kerr solution
in Boyer-Lindquist coordinates. 

The Euclidean section of the boosted black brane in~\eqref{5dmetric4} is then 
\[
ds_{(5)E}^2 = \frac{l^2}{r^2} \frac{dr^2}{W} + 
\frac{r^2}{l^2} W (u_t \, d\tau + \beta dz )^2 
+ \frac{r^2}{l^2} \left( (- \beta \, d\tau + u_t \, dz)^2 + dx^2 + dy^2 
\right) \;.
\]
We now explore the near horizon geometry by adapting a similar
calculation used to examine the Kerr-Newman solution 
in \cite{Mann:1996bi}.
Introducing the new radial variable $R$ by $R^2 = r-r_+$,
the function $W$ has the expansion 
\[
W = \frac{4}{r_+} R^2 + \cdots\;,
\]
around the horizon. Expanding up to order $R^2$, the metric
takes the form
\[
ds^2_{(5)\mathrm{E,NH}} = \frac{l^2}{r_+} \left( 1 -  \frac{R^2}{r_+}\right) 
dR^2 +
\frac{4 r_+}{l^2} R^2 d\chi^2  + 
\frac{r_+^2 + 2 r_+ R^2}{l^2} ( d\tilde{z}^2 + dx^2 + dy^2 ) \;,
\]
where we have replaced the coordinates $\tau$ and $z$ by the new coordinates
\[
\chi = u_t \tau + \beta z \;,\;\;\;
\tilde{z} = u_t z - \beta \tau \;.
\]
We remark that, in contradistinction to the Kerr-Newman solution 
discussed in  \cite{Mann:1996bi}, (i) the coordinate $\tilde{z}$ is
linear rather than angular, i.e.\ we do not need to impose an
identification on it; and (ii) the coordinate $\chi$ is well defined,
since $u_t$ and $\beta$ are constant, so that $u_t d\tau + \beta dz$
is exact. The horizon is at $R=0$. 
The coordinates $x,y,\tilde{z}$
parametrize a three-dimensional plane with a metric which is flat
up to corrections of order $R^2$. This part of the metric is
clearly regular for $R\rightarrow 0$. The variables $R$ and 
$\chi$ parametrize a surface with metric
\[
ds^2_{\rm Cone} = \frac{l^2}{r_+} 
\left( \left[ 1 -  \frac{R^2}{r_+} \right] dR^2 
+ 4 R^2 \frac{r_+^2}{l^4} d\chi^2 \right),
\]
which is, up to a subleading term of order $R^2$, the metric
of a cone with apex at $R=0$. Thus $\chi$ is an angular 
variable and the surface parametrized by $R$ and $\chi$ is
topologically a disk. Imposing the absence of a conical singularity
at $R=0$ fixes the periodicity of $\chi$ to be 
\[
\chi \simeq \chi + 2 \pi \frac{l^2}{2 r_+}  \;.
\]
Since the coordinate $\tilde{z}$ is linear (has no identifications)
we can determine the periodicities of $\tau$ and $z$ from
\[
(\chi, \tilde{z}) \simeq \left( \chi + 2\pi \frac{l^2}{2 r_+}  , \tilde{z}
\right)
\Leftrightarrow (\tau, z) \simeq (\tau + A , z + B)  \;,
\]
with
\[
A = 2 \pi \, u_t\, \frac{l^2}{2r_+}, \qquad
B=2\pi \,\beta\,\frac{l^2}{2r_+}.
\]
The Hawking temperature $T$ is read off from the periodicity of
$\tau$ by $\tau \simeq \tau + T^{-1}$, 
so that
\[
\pi T = \frac{r_+}{l^2 u_t }\;,
\]
which agrees with the result found by computing the surface 
gravity \eqref{Ex:temperature}.

To interpret the periodicity of $z$, remember that the 
boost velocity at the horizon is 
\[
w = - \frac{u_z}{u_t} = - i \frac{\beta}{u_t} \;.
\]
Thus
\[
B = i w \frac{1}{T}\;,
\]
so that the identifications take the form
\[
(\tau, z) \simeq \left( \tau + T^{-1}, z + i w T^{-1} \right) ,
\]
which is analogous to the identification for the Euclidean Kerr
solution, see for example \cite{Mann:1996bi}.

\section{Five-dimensional tidal forces}\label{App:5dtidalforces}
In this appendix we shall construct the frame fields describing the PPON associated to an observer freely falling towards the five-dimensional extremal black brane in \eqref{ExtMetric}. The frame-dragging effects associated to the brane's boost in the $z$ direction mean that an observer who starts falling radially inward from infinity will acquire a velocity in the $z$ direction. We want to pick our first frame field to be the vector field generating the geodesic motion of the observer. To do this, we follow the procedure of~\cite{Copsey:2012gw, Horowitz:2011gh, Lei:2013apa} and introduce the frame field
\begin{equation}\label{e0}
\left( \hat{e}_0 \right)^\mu = \left( \frac{d}{d \tau} \right)^\mu = \dot{t} \left( \partial_t \right)^\mu + \dot{z} \left( \partial_z \right)^\mu + \dot{r} \left( \partial_r \right)^\mu ,
\end{equation}
where $\tau$ is the proper time of our observer or, equivalently, the affine parameter for the geodesic motion, and a  dot denotes differentiation with respect to $\tau$.
 Note that for simplicity we consider an observer who is not moving in the $x$ and $y$ directions.

It is clear that to obtain $\hat{e}_0$, we must first obtain $\dot{t}, \dot{z}$ and $\dot{r}$. To do this, we recall that associated to each of the Killing vector fields $\partial_t, \partial_z, \partial_x, \partial_y$ of \eqref{ExtMetric} there is an integral of motion. These conserved quantities are the energy and momenta,
\begin{align}
E &= - g_{t \mu}  \dot{x}^\mu = \left( \frac{r^2}{l^2} - \frac{\Delta}{r^2l^2} \right) \dot{t} - \frac{\Delta}{r^2l^2} \dot{z}\;, \label{5dE}
\\ p_z &= g_{z \mu} \dot{x}^\mu = \left( \frac{r^2}{l^2} + \frac{\Delta}{r^2l^2} \right) \dot{z} + \frac{\Delta}{r^2l^2} \dot{t}\;, \label{5dPz}
\\ p_x &= g_{x \mu} \dot{x}^\mu = \frac{r^2}{l^2} \dot{x} = 0 \;,
\\ p_y &= g_{y \mu} \dot{x}^\mu = \frac{r^2}{l^2} \dot{y} = 0 \;.
\end{align}
Defining the  quantities
\[
\alpha := \frac{r^2}{l^2} + \frac{\Delta}{r^2l^2}, \quad \beta := \frac{r^2}{l^2} - \frac{\Delta}{r^2l^2}, \quad \gamma := \frac{\Delta}{r^2l^2},
\]
we can simultaneously solve~\eqref{5dE} and~\eqref{5dPz} to find
\begin{align}
\dot{t} &= \frac{l^4}{r^4} \left(\alpha E + \gamma p_z \right) ,
\\ \dot{z} &= \frac{l^4}{r^4} \left( \beta p_z -  \gamma E \right).
\end{align}
Notice that both of these velocities diverge as we approach the horizon at $r_+=0$. This divergence tells us that this particular coordinate system is not valid beyond the horizon. However, for our current purposes, this is not a problem as we are only interested in tidal forces close to, but outside, the horizon. 
In order to write down $\hat{e}_0$, we still need to obtain $\dot{r}$. For this we use that $g_{\mu \nu} \dot{x}^\mu \dot{x}^\nu = -1$ for a timelike observer, which is equivalent to 
\begin{equation}\label{dotr5d}
\dot{r} = - \sqrt{ -\frac{r^2}{l^2}  + \frac{l^2A}{r^2} \left(E-V_+ \right) \left( E-V_- \right)}\;,
\end{equation}
where we've taken the negative root to represent a radially infalling observer and $V_\pm = \frac{1}{\alpha} \left( -\gamma p_z \pm \frac{r^2 p_z}{l^2} \right)$ are the roots of $\alpha E^2 + 2\gamma Ep_z - \beta p_z^2=0$. Notice that had we instead picked the positive root in~\eqref{dotr5d}, describing an outgoing timelike geodesic, $\dot{r}$ will become complex for sufficiently large $r$; this indicates that geodesic cannot reach the boundary but in fact hits a turning point and returns to the bulk~\cite{Maldacena:2011ut, moschella, GibbonsDGnotes}. For this reason, we will only be interested in near horizon tidal forces.

We can now substitute the above expressions for $\dot{t}, \dot{z}, \dot{r}$ into~\eqref{e0} to obtain the following expression for the first frame field
\begin{equation}
\left( \hat{e}_0 \right)^\mu = \frac{l^4}{r^4} \left( \alpha E + \gamma p_z \right) \left( \partial_t \right)^\mu + \frac{l^4}{r^4} \left( \beta p_z - \gamma E \right) \left( \partial_z \right)^\mu - \sqrt{ -\frac{r^2}{l^2}  +\alpha \frac{l^2}{r^2} \left(E-V_+ \right) \left( E-V_- \right)} \left( \partial_r \right)^\mu .
\end{equation}
Whilst the frame field $\hat{e}_0$ correctly describes the parallel propagation, it is not correctly normalised. To form an orthonormal basis of frame fields we can apply Gram-Schmidt procedure to the set of linearly independent frame fields $\hat{e}_a=\{ \hat{e}_0, \hat{e}_1=\partial_r, \hat{e}_2 =\partial_z, \hat{e}_i=\partial_i \}$ where $i=x,y$. This was done using Maple and returns a basis of frame fields that we shall denote $\{ e_a \}$ without the hat. These still correctly characterise the parallel propagation but at the same time are fully orthonormal in the sense that they satisfy $g_{\mu \nu} \left( e_a \right)^\mu \left( e_b \right)^\nu = \eta_{ab}$.

The full expressions for the individual frame fields $\{ e_a \}$ are quite complicated and not especially illuminating so we omit them here. However, we can then use the frame fields as transformation matrices to obtain the components of the Riemann tensor as measured in the PPON via
\begin{equation}\label{RiemannPPON}
\tilde{R}_{abcd} = R_{\mu \nu \rho \sigma} \left( e_a \right)^\mu \left( e_b \right)^\nu \left( e_c \right)^\rho \left( e_d \right)^\sigma .
\end{equation}
The non-zero components of the PPON Riemann tensor are again rather complicated and so rather than provide full expressions, we instead list their scaling behaviour in the near horizon regime in Table \ref{riemanntable}.

\begin{table}
\begin{center}
\begin{tabular}{|c|c|}
\hline Component  & Near horizon behaviour   \\ \hline \rule{0pt}{2.5ex}  $\tilde{R}_{0101}$  & const \\ \hline \rule{0pt}{2.5ex} $\tilde{R}_{0102}$ & $r^{-13}$   \\ \hline \rule{0pt}{2.5ex} 
$\tilde{R}_{0112}$ & $r^{-13}$   \\ \hline \rule{0pt}{2.5ex}
$\tilde{R}_{0202}$ & $r^{-13}$   \\ \hline \rule{0pt}{2.5ex}
$\tilde{R}_{0212}$ & $r^{-13}$   \\ \hline \rule{0pt}{2.5ex}
$\tilde{R}_{0i0j}$ & $\delta_{ij} r^{-6}$   \\ \hline \rule{0pt}{2.5ex}
$\tilde{R}_{0i1j}$ & $\delta_{ij} r^{-6}$   \\ \hline \rule{0pt}{2.5ex}
$\tilde{R}_{0i2j}$ & $\delta_{ij}r^{-6}$  \\ \hline \rule{0pt}{2.5ex}
$\tilde{R}_{1212}$ & $r^{-13}$   \\ \hline \rule{0pt}{2.5ex}
$\tilde{R}_{1i1j}$ & $\delta_{ij}r^{-6}$   \\ \hline \rule{0pt}{2.5ex}
$\tilde{R}_{1i2j}$ & $\delta_{ij}r^{-6}$   \\ \hline \rule{0pt}{2.5ex}
$\tilde{R}_{2i2j}$ & $\delta_{ij}r^{-6}$   \\ \hline  \rule{0pt}{2.5ex} 
$\tilde{R}_{ijkl}$ & $r^{-3} \left( \delta_{il} \delta_{jk} - \delta{ik} \delta_{jl} \right)$  \\ \hline 
\end{tabular}
\end{center}
\caption{Near horizon scaling behaviour of the non-zero components of the five-dimensional Riemann tensor, $\tilde{R}_{abcd}$, as measured in the PPON.}
\label{riemanntable}
\end{table}

\section{Four-dimensional tidal forces}\label{App:4dtidalforces}
To investigate the tidal forces present for the four-dimensional extremal Nernst brane solutions of \cite{Dempster:2015xqa} we must treat the cases with finite and infinite four-dimensional chemical potential separately as we have done throughout the paper. 
These have metrics given in \eqref{4dmetric} and \eqref{4dreducedNernstBraneA=0} respectively. 
We shall proceed in a similar fashion to Appendix \ref{App:5dtidalforces} except for the assumption that the infalling observer is now moving only in the radial direction and has no transverse momentum in either the $x$ or $y$ directions. This is slightly different to the analysis of \cite{Copsey:2012gw} and means the tangent vector for the timelike geodesic on which our radially infalling observer is travelling is given by 
\[
T^\mu = \left( \dot{t}, \dot{r} , \vec{0} \right),
\]
where dot denotes differentiation with respect to  the observer's proper time, $\tau$.

\subsection{$A>0$ tidal forces}

The extremal version of \eqref{4dmetric} is given by
\begin{equation}\label{4dAneq0extmetric}
ds^2_{A>0, \text{ Ext}} = \frac{r}{l} \left( - \frac{r^2}{l^2 \left( 1 + \frac{\Delta}{A r^4} \right)^{1/2}} dt^2 + \frac{l^2 \left( 1 + \frac{\Delta}{A r^4} \right)^{1/2}}{r^2} dr^2 + \frac{r^2}{l^2}  \left( 1 + \frac{\Delta}{A r^4} \right)^{1/2} \left( dx^2 + dy^2 \right) \right) .
\end{equation}
The energy is again an integral of motion:
\[
E=-g_{tt} \dot{t} = \frac{r^3}{l^3 \left( 1 + \frac{\Delta}{A r^4} \right)^{1/2}} \dot{t} \quad \Rightarrow \quad
 \dot{t} = \frac{l^3 E \left( 1 + \frac{\Delta}{A r^4} \right)^{1/2}}{r^3} .
\]
For a timelike geodesic we have
\[ 
g_{\mu \nu} T^\mu T^\nu = -1 \quad \Rightarrow \quad
\dot{r} = - \frac{1}{l^{1/2}r} \sqrt{l^3E^2- \frac{r^3}{\left( 1 + \frac{\Delta}{A r^4} \right)^{1/2}}},
\]
where we pick the negative square root to represent an observer falling radially inwards. We could equally well pick the positive root and consider an outgoing geodesic but $\dot{r}$ will become complex for large $r$, meaning the geodesic encounters a turning point and is reflected back into the bulk. This is reminiscent of the situation in Appendix~\ref{App:5dtidalforces} and in fact, this inability of timelike geodesics to reach the boundary is an example of a property that hyperscaling-violating Lifshitz spacetimes can inherit from their parent Anti de-Sitter spacetimes. All of this means that we need only focus on the ingoing observer and near horizon tidal forces. Another similarity with Appendix~\ref{App:5dtidalforces} is the divergence of $\dot{t}$ and $\dot{r}$ as $r \rightarrow 0$; again this indicates the coordinates are only valid up the horizon which is absolutely fine for the analysis of tidal forces.

Next we align the frame field\footnote{We use unhatted frame fields in four-dimensions to distinguish from their hatted cousins in five-dimensions.} $e_0$ with the vector field $\frac{d}{d \tau}$ responsible for generating the integral curve along which the observer is moving: 
\begin{align*}
(e_0)^\mu = \left( \frac{d}{d \tau} \right)^\mu &= \dot{t} \partial_t^\mu + \dot{r} \partial_{r}^\mu 
\\ 
&= \frac{l^3 E \left( 1 + \frac{\Delta}{A r^4} \right)^{1/2}}{r^3} \partial_t^\mu - \frac{1}{l^{1/2}r} \sqrt{l^3E^2- \frac{r^3}{\left( 1 + \frac{\Delta}{A r^4} \right)^{1/2}}} \partial_{r}^\mu .
\end{align*}
The observer is moving in the $(t,r)$ directions and so there are two frame fields associated to this: $e_0$ and $e_1$. 
Since the observer isn't moving in any of the $x^i$ ($i \geq 2$) directions, the frames $e_i$ for $i \geq 2$ are just given by the square roots of the inverse metric components i.e.
\[ 
(e_i)^\mu = \frac{l}{r \left( 1 + \frac{\Delta}{A r^4} \right)^{1/4}} \partial_i^\mu  .
\]
It remains to find the frame $e_1$ such that the $\{ e_a \}$ form a PPON.
 We have picked $e_0$ to describe the parallel propagation and so we just need a second frame field, $e_1$, that is orthonormal to both $e_0$ and $e_i, i \geq 2$. It follows from simple linear algebra that
\[
(e_1)^\mu = - \frac{l^{3/2} \left( 1 + \frac{\Delta}{A r^4} \right)^{1/2}}{r^3} 
\sqrt{l^3E^2-\frac{r^3}{\left( 1 + \frac{\Delta}{A r^4} \right)^{1/2}}} \partial_t^\mu + \frac{lE}{r} \partial_r^\mu .
\]
It is interesting to note that in the case of the static four-dimensional metric, the frame fields are already orthonormal whereas in Appendix~\ref{App:5dtidalforces}, where the five-dimensional metric is non-static, this is not the case and we had to perform an additional Gram-Schmidt procedure at this point.

We next use Maple to find the components of the Riemann tensor in a coordinate basis with lowered indices, $R_{\mu \nu \rho \sigma}$, and then multiply by frame fields to obtain the local tidal forces felt by the observer as in \eqref{RiemannPPON}. We again omit the full expressions and instead list in Table~\ref{4dAneq0tidalforces} the scaling behaviour of the non-zero components in the near horizon regime
\begin{table}
\begin{center}
\begin{tabular}{|c|c|}
\hline Component  & Near horizon behaviour   \\ \hline \rule{0pt}{2.5ex}  $\tilde{R}_{0101}$  & $r$ \\ \hline \rule{0pt}{2.5ex} $\tilde{R}_{0i0j}$ & $\delta_{ij} r^{-4}$   \\ \hline \rule{0pt}{2.5ex} 
$\tilde{R}_{0i1j}$ & $\delta_{ij} r^{-4}$   \\ \hline \rule{0pt}{2.5ex}
$\tilde{R}_{1i1j}$ & $\delta_{ij} r^{-4}$   \\ \hline \rule{0pt}{2.5ex}
$\tilde{R}_{ijkl}$ & $r \left( \delta_{il} \delta_{jk} - \delta{ik} \delta_{jl} \right)$  \\ \hline 
\end{tabular}
\end{center}
\caption{Near horizon scaling behaviour of the non-zero components of the four-dimensional $A>0$ Riemann tensor, $\tilde{R}_{abcd}$, as measured in the PPON.}
\label{4dAneq0tidalforces}
\end{table}

\subsection{$A=0$ tidal forces}
Here we repeat the same procedure as above for the $A=0$ extremal metric. The extremal version of~\eqref{4dreducedNernstBraneA=0} is given by
\begin{equation}\label{4dA=0extmetric}
ds_{A=0, \text{ Ext}}^2 = - \frac{r^5}{\Delta^{1/2}l^3} dt^2 + \frac{\Delta^{1/2} l}{r^3} dr^2 + \frac{\Delta^{1/2}r}{l^3} \left( dx^2 + dy^2 \right).
\end{equation}
The resulting nonzero components of the Riemann tensor as measured in the PPON are given in Table~\ref{4dA=0tidalforces}.
\begin{table}
\begin{center}
\begin{tabular}{|c|c|}
\hline Component  & Near horizon behaviour   \\ \hline \rule{0pt}{2.5ex}  $\tilde{R}_{0101}$  & $r$ \\ \hline \rule{0pt}{2.5ex} $\tilde{R}_{0i0j}$ & $\delta_{ij} r^{-4}$   \\ \hline \rule{0pt}{2.5ex} 
$\tilde{R}_{0i1j}$ & $\delta_{ij} r^{-4}$   \\ \hline \rule{0pt}{2.5ex}
$\tilde{R}_{1i1j}$ & $\delta_{ij} r^{-4}$   \\ \hline \rule{0pt}{2.5ex}
$\tilde{R}_{ijkl}$ & $r^3 \left( \delta_{il} \delta_{jk} - \delta{ik} \delta_{jl} \right)$  \\ \hline 
\end{tabular}
\end{center}
\caption{Near horizon scaling behaviour of the non-zero components of the four-dimensional $A=0$ Riemann tensor, $\tilde{R}_{abcd}$, as measured in the PPON.}
\label{4dA=0tidalforces}
\end{table}

\subsection{Consistency with existing classification}
The near horizon scaling behaviours of the PPON Riemann tensor components in Tables~\ref{4dAneq0tidalforces} and~\ref{4dA=0tidalforces} agree. This is consistent with the fact that the parameter $A$ only affects the asymptotic geometry, which is why the metrics~\eqref{4dAneq0extmetric} and~\eqref{4dA=0extmetric} both take the same form in the small $r$ limit: specifically, a hyperscaling-violating Lifshitz metric with parameters $(z,\theta)=(3,1)$ as observed in~\cite{Dempster:2015xqa}. 

It is worthwhile to check the consistency of the results of this appendix with the complete classification of hyperscaling-violating Lifshitz singularities obtained in~\cite{Copsey:2012gw}. It can be shown that our $(z,\theta)=(3,1)$ geometry is equivalent to a $(n_0,n_1)=(10,4)$ geometry in their notation. This would place our near horizon metric into Class IV of the analysis in~\cite{Copsey:2012gw}, making it both consistent with the Null Energy Condition and indicative of a null curvature singularity (infinite tidal forces) at $r=0$.

\section{Normalization of the vector potential \label{Vanishing}}

For the four-dimensional chemical potential
$\mu \sim A_t(r=\infty)$, to be uniquely defined, 
it is crucial that the vector potential is normalized
such that $A_t(r_+)=0$. While this is widely used and
the reason well known, see for example 
\cite{Kobayashi:2006sb,Hartnoll:2009sz}, we would like to 
review the full argument here for completeness. 

Assume that we are given a static space-time which has a Killing
horizon with Killing vector field $\xi$. If the norm of $\xi$ has a simple
zero at the horizon, in other words, if the solution is non-extremal,
then the space-time can be continued analytically to a space-time
which contains a bifurcate horizon \cite{Racz:1995nh}. This means
that the horizon has a spatial section $\Sigma_0$ where the Killing vector
field $\xi$ vanishes. If $A$ is a well-defined one-form on this space-time,
then $A(\xi)=0$ on $\Sigma_0$. Since the horizon is generated by the
flow of the Killing vector field $\xi$, and if assuming that the 
one-form $A$ is invariant under $\xi$, $L_\xi A =0$ (where $L_\xi$ denotes the
Lie derivative), 
it follows that $A(\xi)=0$
on the whole horizon. Outside the horizon we can define a time 
coordinate $t$, such that $\xi =\partial_t$. Then the horizon limit
of the component $A_t$ of the one-form is $A_t \rightarrow A(\xi) = 0$. 

In our application, we have non-extremal solutions with Killing 
horizons, generated by $\xi$, given by 
$\xi=\partial_t$ outside the horizon. Moreover not only the metric but 
also the 
vector field is assumed static (invariant under $t$), and therefore $A_t$
has to vanish on the horizon. By continuity this continues to hold
in the extremal limit.
\newpage

\providecommand{\href}[2]{#2}\begingroup\raggedright\endgroup

\end{document}